\documentclass[sigplan,10pt]{acmart}
\AtBeginDocument{%
  }

\renewcommand\footnotetextcopyrightpermission[1]{}
\usepackage{tikz}
\usepackage{booktabs}
\usepackage{pgfplots}
\usepackage{ifthen}
\usepackage{xspace}
\usetikzlibrary{patterns,fit,positioning,arrows.meta,calc}
\usepgfplotslibrary{groupplots}
\pgfplotsset{compat=1.18}

\newboolean{revisionMode}
\setboolean{revisionMode}{true}

\newcommand{\SystemName}{\texttt{EEP}\xspace}

\newcommand{\revision}[1]{
  \unskip\ifthenelse{\boolean{revisionMode}}
    {\textcolor{brown}{#1}}
    {#1}\unskip
}

\begin{document}


\title{\bf Surviving Partial Rank Failures in \\Wide Expert-Parallel MoE Inference}

\author{
Xun Sun$^{1,2}$, 
Shaoyuan Chen$^{1}$, 
Pingchuan Ma$^{1}$, 
Yue Chen$^{3}$, 
Ziwei Yuan$^{1,5}$, 
Zhanhao Cao$^{3}$, 
Han Han$^{2}$, 
Shangming Cai$^{4}$, 
Teng Ma$^{4}$, 
Xuchun Shang$^{4}$, 
Xinpeng Zhao$^{4}$, 
Ke Yang$^{3}$, 
Junlin Wei$^{5}$, 
Lianzhi Lin$^{5}$, 
Yuji Liu$^{5}$, 
Feng Ren$^{1}$, 
Haoran Hu$^{1}$, 
Cheng Wan$^{6}$, 
Yingdi Shan$^{1}$, 
Yongwei Wu$^{1}$, 
Mingxing Zhang$^{1,}$\footnotemark[1]
}

\affiliation{
\institution{
$^{1}$Tsinghua University \quad
$^{2}$ByteDance \quad
$^{3}$Approaching AI \quad
$^{4}$Alibaba Cloud Computing \\
$^{5}$JD.com \quad
$^{6}$Georgia Institute of Technology
}
\country{}
}

\begin{abstract}

Mixture-of-Experts (MoE) serving relies on wide expert parallelism (EP) to aggregate the memory capacity and bandwidth of many GPUs within one inference instance. This efficiency comes with a systems cost: every decoding step depends on token dispatch and combination across all active EP ranks, so even one rank failure can disrupt the entire service. Existing EP stacks handle such failures poorly because they treat membership as a fixed configuration established at initialization. The same rank set determines communicator state, expert placement, and the routing metadata baked into CUDA execution graphs, leaving the system with no way to shrink around a failure while keeping the instance valid.

This paper argues that partial-failure tolerance should instead be formulated as a live EP validity problem. We present \SystemName, a communication and runtime substrate that represents membership as explicit, mutable runtime state. \SystemName repairs the specific state invalidated by a fault: it restores peer reachability without rebuilding the communication substrate, repairs lost expert coverage through a bandwidth-aware hierarchy, and reintegrates repaired ranks without forcing healthy ranks to recapture their CUDA graphs.

We implement \SystemName in an EP serving stack integrated with SGLang and evaluate it under steady-state serving, failure recovery, and rank reintegration. The results show that explicit mutable membership preserves the steady-state fast path, staying within 4.4\% of a fixed-membership DeepEP baseline under static serving, while turning a local rank fault from whole-instance downtime into two bounded interruptions. On a single-rank failure workload, \SystemName incurs an 11\,s recovery pause and an 8\,s reintegration pause, and restores throughput to within 95\% of the pre-fault level within 52\,s, whereas a fixed-membership full-restart baseline remains unavailable until 348\,s.

\end{abstract}



\keywords{Mixture-of-Experts, expert parallelism, LLM serving, fault tolerance, GPUDirect RDMA}

 \maketitle

\footnotetext[1]{Correspond to: \href{mailto:zhang\_mingxing@mail.tsinghua.edu.cn}{zhang\_mingxing@mail.tsinghua.edu.cn}}
\thispagestyle{plain}
\pagestyle{plain}

\section{Introduction}
\label{sec:introduction}

Wide expert parallelism (EP) has become the standard strategy for serving large Mixture-of-Experts (MoE) models at production scale: dozens of GPUs cooperate on every decoding step, aggregating expert weights, memory capacity, and memory bandwidth within a single EP instance~\cite{deepseekai2024deepseekv3technicalreport,vllm_ep_docs,nccl_ep,deepseekmoe,deepseekv2}. At the scale where EP is necessary, however, partial failures are routine. GPU, process, and network faults occur regularly in large clusters --- in one production datacenter deployment, GPU and node failures accounted for over 78\% of hardware fault events at a rate of 382 events per month~\cite{tent2026}. Studies of large-scale ML clusters~\cite{hpca-reliability} further confirm that as job size increases, failure rates grow significantly, making partial failures a steady operational condition rather than an edge case --- yet existing EP stacks cannot survive them without restarting the entire serving instance from scratch.

The consequences are disproportionate. Large MoE EP instances are long-lived, latency-sensitive services. When one rank in a 32-GPU EP instance fails, the desirable behavior is clear: shrink around the failure, restore a valid reduced-capacity configuration quickly, and recover the lost capacity when the rank returns. Instead, existing systems rebuild the entire instance, turning a localized hardware event into a multi-minute outage.

\textbf{The root cause is that existing EP systems encode membership as fixed structure established at initialization.} The same rank set determines communicator state, expert placement, and the routing metadata baked into CUDA execution graphs. A single-rank failure therefore does not remove only one worker. It invalidates the peer set that communication should target, may eliminate logical experts held on the failed device, and can make the graph-captured execution state inconsistent with the running instance. Existing stacks have no way to repair these states independently, so a local fault becomes a whole-instance rebuild.

This coupling creates three tensions that make partial-failure tolerance hard to engineer.

\paragraph{Tension 1: CUDA-graph execution and online reconfiguration are structurally opposed.}
CUDA graphs are essential for high-performance serving because they eliminate per-step kernel launch overhead, but they do so by freezing kernel structure, launch parameters, and pointer identities at capture time~\cite{cuda_graphs_blog,cuda_programming_guide}. Membership changes require the effective peer set and routing state to change after a failure or repair. The straightforward approaches all share the same deficiency: shrinking or recreating a communicator still leaves the EP state above it to be rebuilt, while recapturing CUDA graphs on healthy ranks makes recovery cost scale with model size and system scale. Any solution must therefore keep graph structure intact on healthy ranks while still changing the peers they communicate with and the routing they apply.

\paragraph{Tension 2: fast degraded-mode recovery and MoE semantic completeness are structurally opposed.}
After a rank fails, the service should resume quickly on the surviving ranks. But MoE execution cannot simply omit the experts that disappeared with the failed device: once a logical expert becomes unreachable, subsequent requests are no longer executing the intended model. Neither alternative is acceptable: continuing immediately over the reduced peer set risks semantic incompleteness, while waiting for the failed rank to restart or reloading all missing experts from storage collapses availability. Recovery must therefore restore expert coverage over the surviving ranks before the system can safely continue, without waiting for the failed rank.

\paragraph{Tension 3: asynchronous rank recovery and synchronous distributed-runtime initialization are structurally opposed.}
Even after a valid reduced-capacity instance has been restored, throughput remains below baseline until the failed rank is repaired and rejoined. Existing distributed runtimes, however, assume that initialization, process-group formation, metadata exchange, and CUDA-graph capture occur with the whole group participating together. If the recovering rank follows that path, healthy ranks are drawn into collective barriers, model initialization and graph recapture, costing tens of seconds, turning reintegration into a long service interruption. Reintegration must therefore let the recovering rank complete its initialization in isolation without forcing healthy ranks off the serving critical path.

These tensions suggest a different formulation of partial-failure tolerance for MoE serving. The goal is not to make restart faster. \textbf{The goal is to keep a live EP instance valid across membership changes.} An EP instance is valid only if communication targets active peers, every logical expert is reachable on at least one active rank, and CUDA-graph-captured dispatch and combination continue to execute against routing state that matches the current membership. The systems question is how to represent and repair these three states without rebuilding the instance.

We present \SystemName, built around one principle: represent membership as explicit, mutable runtime state rather than fixed communication structure. \SystemName is a communication and runtime substrate for elastic expert-parallel inference. Rather than rebuilding the EP instance after each membership change, it repairs the specific state invalidated by the fault: peer reachability, expert coverage, and graph-visible routing. It restores expert coverage after failures through a three-tier hierarchy---local reuse, GPU-to-GPU relocation, and DRAM-backed reload. It also keeps communication and routing state graph-stable but mutable at runtime, so healthy ranks continue replaying the same captured graphs across failures and rejoins. Together, these mechanisms change the unit of recovery from ``rebuild the EP instance'' to ``repair the invalidated peer set, expert coverage, and routing state.''

\begin{figure}[t]
    \centering
    \includegraphics[width=\columnwidth]{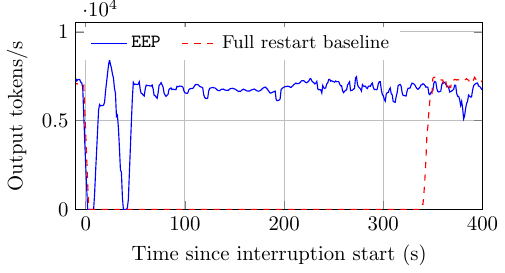}
    \caption{Throughput after a single-rank failure. \SystemName incurs two bounded pauses, one for recovery and one for reintegration, while the fixed-membership full-restart baseline remains unavailable for several minutes because it repeats the full initialization path before serving resumes.}
    \label{fig:intro-reintegration-hero}
\end{figure}

Figure~\ref{fig:intro-reintegration-hero} illustrates the effect on a live EP instance that loses one rank. In a fixed-membership design, the loss forces a full restart: a single uninterrupted outage lasting 348\,s on this workload, because the system must redo the initialization path, including environment setup, model loading, JIT warmup, and CUDA-graph capture. \SystemName instead produces two bounded interruption windows---an 11\,s recovery pause followed by a productive reduced-capacity serving plateau, then an 8\,s reintegration pause as the repaired rank rejoins---for about 19\,s of total off-service time. The central claim of this paper is that a local rank fault need not become whole-instance downtime.

We implement \SystemName in an EP serving stack integrated with SGLang and evaluate it under steady-state serving, partial-failure recovery, and rank reintegration. The evaluation answers three questions: what the overhead of explicit mutable membership is relative to a fixed-membership baseline under static serving, how quickly a valid reduced-capacity instance can be restored after failure compared to full-instance restart, and whether repaired ranks can rejoin without forcing healthy-rank graph recapture.

In summary, this paper makes the following contributions:

\begin{itemize}
    \item \textbf{A formulation of live EP validity under partial failures.}

    We show that partial-failure tolerance in wide EP is a state-validity problem, not a faster-restart problem, and define a live EP instance as valid iff three conditions hold simultaneously: every peer-table entry points to a reachable rank, every logical expert is hosted by at least one active rank, and CUDA-graph-captured dispatch and combination route tokens only to valid experts on active ranks. This formulation makes the recovery contract precise and checkable, and clarifies why peer-set repair, expert-coverage repair, and graph-visible-state repair are each necessary.

    \item \textbf{A design for repairing EP instances without rebuilding them.}

    We design \SystemName around explicit mutable membership, using a GPU-resident peer table for communication continuity, a three-tier expert-repair hierarchy for restoring lost coverage, and graph-stable peer and routing tables with a deferred-join protocol to reintegrate repaired ranks without healthy-rank graph recapture or global synchronization barriers.

    \item \textbf{An end-to-end implementation and evaluation in an MoE serving stack.}

    We integrate \SystemName into SGLang and show that it stays within 4.4\% of a fixed-membership DeepEP baseline under static serving, and recovers after bounded 6--21\,s interruption windows for 1-, 2-, 4-, 8-, and 16-rank failures, while the fixed-membership full restart remains unavailable until 348\,s on the same workload.
\end{itemize}
\begin{table*}[t]
\centering
\small
\begin{tabular}{l p{5.5cm} p{3.6cm} p{4.6cm}}
\toprule
\textbf{Library} &
\textbf{Behavior when a process fails} &
\textbf{Support for membership reconfiguration} &
\textbf{Notes} \\
\midrule

\textbf{Gloo} &
Reports an error and invalidates the entire process group; all remaining workers abort. &
Not supported (world size fixed at initialization). &
Used for CPU collectives and metadata synchronization in PyTorch. \\

\textbf{NCCL} &
Collectives on a faulted communicator may error or stall; recent NCCL versions add health reporting and allow applications to shrink or recreate communicators after failures. &
Supported only via communicator shrink or re-initialization; membership remains a communicator property. &
High-performance GPU collectives with improving fault-handling support, but EP state above NCCL must still adapt to the new communicator. \\

\textbf{DeepEP} &
Optimizes MoE dispatch/combine and can surface communication failures, but dispatch/combine still execute over a preconfigured EP group. &
Not supported as live EP membership repair. &
Specialized EP communication library; does not by itself repair expert placement or reintegrate ranks. \\

\bottomrule
\end{tabular}
\caption{Behavior of existing communication libraries under failure and membership change. From the serving runtime's perspective, they still treat membership as communicator or EP-group state rather than mutable runtime state.}
\label{tab:comm-behavior}
\end{table*}

\section{Background and Motivation}

\subsection{Large-Scale Expert Parallelism}

\begin{figure}[t]
    \centering
    \includegraphics[width=0.9\linewidth]{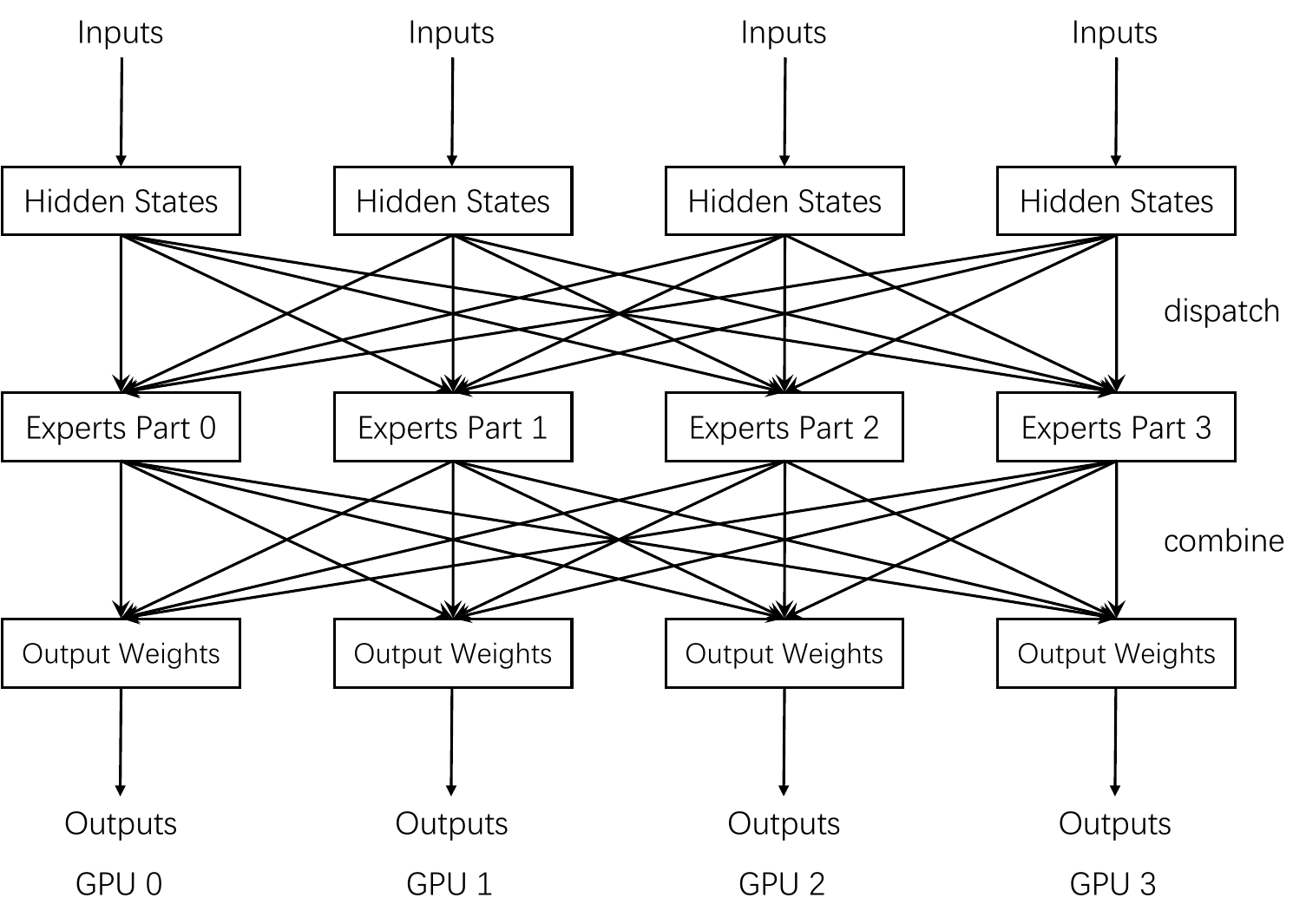}
    \caption{Expert parallelism routes tokens to remote experts and gathers the results back. Conventional implementations bind this pattern to a fixed communicator; \SystemName executes it over a dynamic peer set.}
    \label{fig:ep}
\end{figure}

Mixture-of-Experts (MoE) models activate only a small subset of experts per token, enabling large parameter counts with substantially lower per-token FLOPs than dense architectures~\cite{shazeer2017moe,switchtransformers,deepseekmoe,deepseekv2}. During inference, a router selects experts for each token; the system then dispatches token activations to the devices hosting those experts and collects the results. To scale this across many GPUs, contemporary systems use expert parallelism (EP): experts are partitioned across devices, and each inference step triggers two communication phases---token dispatch and token combination---as illustrated in Figure~\ref{fig:ep}.

The communication pattern in Figure~\ref{fig:ep} sits directly on the decoding critical path: every forward pass depends on dispatch and combination completing across the full EP group. When EP scales to dozens of GPUs, this creates a hard dependency on every participating rank. If one rank disappears, some experts become unreachable and the communication pattern expected by the surviving ranks no longer matches reality.

The deeper issue is that most EP stacks do not represent membership as mutable runtime state. Table~\ref{tab:comm-behavior} summarizes how existing communication libraries handle failures. General-purpose libraries such as NCCL and Gloo tie membership to a process group or communicator established at initialization~\cite{gloo}. Specialized libraries such as DeepEP handle token dispatch and combination for a chosen EP group~\cite{deepep}. These libraries differ in how much fault handling they expose, but from the serving runtime's perspective they all treat membership as communicator-bound state rather than something that can be repaired in place. Recent NCCL releases add RAS health support and communicator-shrink workflows~\cite{nccl_ras,nccl_fault_blog,nccl_ep}, which is meaningful progress, but the application must still rebuild communicator-bound EP state around the new rank set after a shrink.

The same static-membership assumption appears above the communication layer. Expert placement is decided during initialization, so a rank failure immediately causes the experts on that device to become unreachable. Router metadata, placement state, and auxiliary synchronization all assume the same rank set. A membership change is therefore not merely a communication error; it is a state-validity failure that simultaneously breaks peer reachability, expert coverage, and routing correctness.

Taken together, these observations reframe the problem: partial-failure tolerance in EP is fundamentally a live EP validity problem rather than a faster-restart problem. The remainder of this section provides the technical background for two hardware features---GPU-initiated RDMA and CUDA execution graphs---whose properties directly shape what solutions are possible.

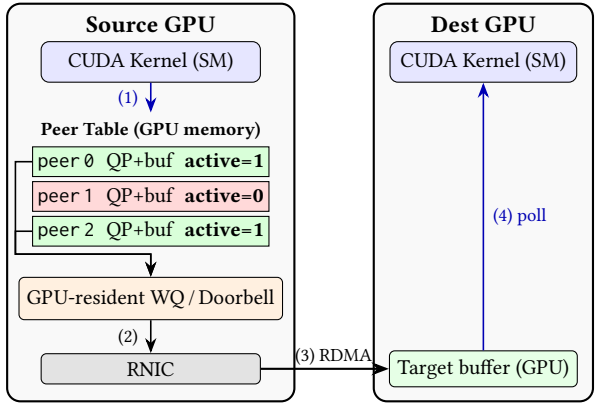
\begin{figure}[t]
\centering
\begin{tikzpicture}[
  font=\footnotesize,
  >=Stealth,
  box/.style={draw, rounded corners=2pt, align=center, inner sep=3pt, fill=white},
  kernel/.style={draw, rounded corners=3pt, align=center, inner sep=4pt,
                 fill=blue!10, minimum width=2.9cm, minimum height=0.48cm},
  ptrow/.style={draw, minimum width=2.9cm, minimum height=0.36cm, align=center, inner sep=2pt},
  active/.style={fill=green!15},
  dead/.style={fill=red!15},
  arr/.style={->, semithick},
  rdma/.style={->, thick, black},
]

\draw[draw=black, thick, rounded corners=5pt, fill=gray!5]
  (-0.15,0.29) rectangle (3.65,5.55);
\node[font=\small\bfseries] at (1.75,5.28) {Source GPU};

\node[kernel] (kern) at (1.75,4.78) {CUDA Kernel (SM)};

\node[font=\scriptsize\bfseries] (table) at (1.75,3.87) {Peer Table (GPU memory)};
\node[ptrow, active] (p0) at (1.75,3.46)
  {\texttt{peer\,0}\;\;QP+buf\;\;\textbf{active=1}};
\node[ptrow, dead]   (p1) at (1.75,3.00)
  {\texttt{peer\,1}\;\;QP+buf\;\;\textbf{active=0}};
\node[ptrow, active] (p2) at (1.75,2.54)
  {\texttt{peer\,2}\;\;QP+buf\;\;\textbf{active=1}};

\node[box, fill=orange!12, minimum width=2.9cm, minimum height=0.56cm]
  (wq) at (1.75,1.65)
  {GPU-resident WQ\,/\,Doorbell};

\node[box, fill=gray!22, minimum width=2.9cm, minimum height=0.40cm]
  (rnic) at (1.75,0.72) {RNIC};

\draw[draw=black, thick, rounded corners=5pt, fill=gray!5]
  (4.70,0.29) rectangle (7.60,5.55);
\node[font=\small\bfseries] at (6.15,5.28) {Dest GPU};

\node[kernel, minimum width=2.4cm] (rkern) at (6.15,4.78) {CUDA Kernel (SM)};
\node[box, fill=green!10, minimum width=2.4cm, minimum height=0.44cm]
  (rbuf) at (6.15,0.72) {Target buffer (GPU)};

\draw[arr, blue!70!black]
  (kern.south) -- (table.north);
\node[font=\scriptsize, text=blue!70!black] at (1.45,4.30) {(1)};

\coordinate (mid1) at ($ (wq.north) + (0,0.3) $);

\draw[arr] (p0.west) -- +(-0.22,0) |- (mid1) -- (wq.north);
\draw[arr] (p2.west) -- +(-0.22,0) |- (mid1) -- (wq.north);

\draw[arr] (wq.south) -- (rnic.north);
\node[font=\scriptsize] at (1.45,1.14) {(2)};

\draw[rdma] (rnic.east) -- +(0.85,0) |- (rbuf.west);
\node[font=\scriptsize, text=black] at (4.17,0.88) {(3) RDMA};

\draw[arr, blue!70!black]
  (rbuf.north) -- (rkern.south)
  node[midway, right, font=\scriptsize] {(4) poll};

\end{tikzpicture}
\caption{GPU-driven EP communication path in \SystemName. Kernels read the peer table on device, issue transfers directly from the GPU, and skip inactive peers without recapturing the CUDA graph.}
\label{fig:ibgda}
\end{figure}

\subsection{GPUDirect RDMA}
\label{sec:gdr}

High-performance MoE inference requires fast, fine-grained GPU-to-GPU communication across machines. Modern clusters use RDMA-capable NICs (RNICs) to exchange data with low latency and minimal CPU involvement. In the traditional programming model, the CPU registers memory, prepares work requests, and rings NIC doorbells for every communication step. This model suits bulk transfers but is poorly matched to EP dispatch and combination: data volumes are small, communication rounds are frequent, and any host involvement falls directly on the decoding critical path~\cite{gpudirect_rdma}.

GPUDirect RDMA addresses this by exposing NIC resources---doorbell registers and work queues---directly into the GPU's address space, so that CUDA kernels can post work, ring the doorbell, and trigger RDMA transfers without CPU involvement~\cite{gpudirect_rdma}. NVSHMEM with IBGDA further demonstrates that this GPU-initiated model can achieve high throughput for fine-grained transfers~\cite{nvshmem-ibgda}, and recent NCCL extensions generalize this capability within NCCL's production infrastructure~\cite{nccl-gin}.

The consequence relevant to this paper is not just lower latency. GPU-initiated RDMA turns EP communication into a sequence of GPU-side point-to-point actions whose targets can be chosen from mutable device-side state at runtime. This is what makes it possible to skip failed peers atomically within a CUDA kernel, continue over a reduced set, and later reactivate repaired peers---all without inserting a CPU-managed membership query back into the dispatch/combination path. Figure~\ref{fig:ibgda} shows the resulting communication path: kernels read the peer table, post work queue entries for active peers only, and ring the doorbell directly from GPU execution. The takeaway is that any solution to Tension~1 must keep membership decisions on-device.

\subsection{CUDA Graphs}

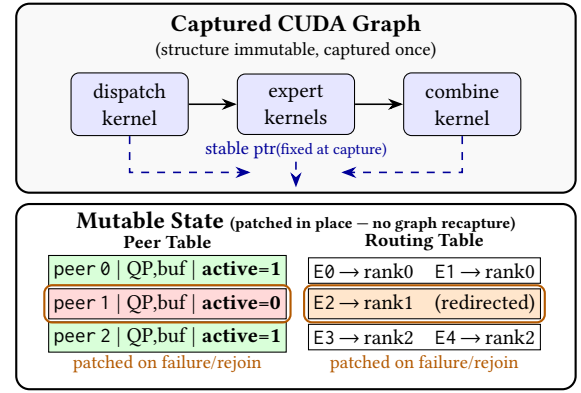
\begin{figure}[t]
\centering
\begin{tikzpicture}[
  font=\footnotesize,
  >=Stealth,
  box/.style={draw, rounded corners=2pt, align=center, inner sep=3pt, fill=white},
  kernel/.style={draw, rounded corners=3pt, align=center, inner sep=4pt,
                 fill=blue!10, minimum width=1.55cm, minimum height=0.48cm},
  ptrow/.style={draw, minimum width=2.6cm, minimum height=0.32cm, align=center,
                inner sep=2pt, fill=white},
  arr/.style={->, semithick},
  ptr/.style={->, semithick, blue!60!black, dashed},
  patch/.style={draw=orange!70!black, thick, rounded corners=3pt},
]

\draw[draw=black, thick, rounded corners=5pt, fill=gray!5]
  (-0.45, 2.35) rectangle (6.95, 4.85);
\node[font=\small\bfseries, align=center] at (3.25, 4.42)
  {Captured CUDA Graph\\[-2pt]{\scriptsize\textnormal{(structure immutable, captured once)}}};

\node[kernel] (dispatch) at (1.05, 3.52) {dispatch\\kernel};
\node[kernel] (expert)   at (3.25, 3.52) {expert\\kernels};
\node[kernel] (combine)  at (5.45, 3.52) {combine\\kernel};

\draw[arr] (dispatch.east) -- (expert.west);
\draw[arr] (expert.east)   -- (combine.west);

\draw[ptr] (dispatch.south) -- +(0,-0.48) -- +(1.6,-0.48);
\draw[ptr] (combine.south)  -- +(0,-0.48) -- +(-1.6,-0.48);
\node[font=\scriptsize, text=blue!60!black, align=center] at (3.25,2.92)
  {stable ptr{\tiny(fixed at capture)}};
\draw[ptr] (3.25,2.76) -- (3.25,2.42);

\draw[draw=black, thick, rounded corners=5pt, fill=white]
  (-0.45, -0.25) rectangle (6.95, 2.20);
\node[font=\small\bfseries] at (3.25, 1.97)
  {Mutable State \tiny{(patched in place --- no graph recapture)}};

\node[font=\scriptsize\bfseries] at (1.55,1.70) {Peer Table};
\node[ptrow, fill=green!15] (pt0) at (1.55,1.32)
  {\texttt{peer\,0}\;|\;QP,buf\;|\;\textbf{active=1}};
\node[ptrow, fill=red!15]   (pt1) at (1.55,0.88)
  {\texttt{peer\,1}\;|\;QP,buf\;|\;\textbf{active=0}};
\node[ptrow, fill=green!15] (pt2) at (1.55,0.44)
  {\texttt{peer\,2}\;|\;QP,buf\;|\;\textbf{active=1}};

\draw[patch] ($(pt1.north west)+(-0.05,0.04)$) rectangle ($(pt1.south east)+(0.05,-0.04)$);
\node[font=\scriptsize, text=orange!70!black] at (1.55,0.08) {patched on failure/rejoin};

\node[font=\scriptsize\bfseries] at (4.95,1.70) {Routing Table};
\node[ptrow] (rt0) at (4.95,1.32)
  {\texttt{E0}\,$\rightarrow$\,rank0\quad\texttt{E1}\,$\rightarrow$\,rank0};
\node[ptrow, fill=orange!20] (rt1) at (4.95,0.88)
  {\texttt{E2}\,$\rightarrow$\,rank1\quad (redirected)};
\node[ptrow] (rt2) at (4.95,0.44)
  {\texttt{E3}\,$\rightarrow$\,rank2\quad\texttt{E4}\,$\rightarrow$\,rank2};

\draw[patch] ($(rt1.north west)+(-0.05,0.04)$) rectangle ($(rt1.south east)+(0.05,-0.04)$);
\node[font=\scriptsize, text=orange!70!black] at (4.95,0.08) {patched on failure/rejoin};

\end{tikzpicture}
\caption{CUDA-graph-stable reconfiguration in \SystemName. The graph keeps stable pointers to peer and routing tables whose contents can be patched in place across failure and reintegration.}
\label{fig:cuda-graph}
\end{figure}

CUDA graphs capture a GPU workload once and replay it with minimal CPU involvement per step, which is essential for the throughput targets of MoE serving~\cite{cuda_graphs_blog,cuda_programming_guide}. The benefit comes with a structural constraint: launch order, kernel sequencing, and pointer identities are fixed at capture time. Only the contents of memory reachable through those fixed pointers may change across replays~\cite{cuda_programming_guide}.

This constraint is the other half of Tension~1. If peer identities, routing metadata, or synchronization state are embedded directly in graph structure, a membership change invalidates the captured graph and forces recapture on the affected ranks. Communicator shrink does not solve this; it only moves the rebuild boundary upward, because the EP state above the communicator still must be rebuilt around the new rank set. In large MoE deployments, graph recapture takes seconds to minutes and becomes part of the failure-recovery latency.

The practical implication is sharp: any mechanism for partial-failure tolerance must preserve graph structure on healthy ranks. Membership changes may patch the contents of graph-visible state, but they cannot require rebuilding the graph itself. This motivates \SystemName's design choice of stable graph-visible pointers to mutable peer and routing tables, shown in Figure~\ref{fig:cuda-graph} and described in Section~\ref{sec:impl-scaling}.
\section{System Design}
\label{sec:design}

\SystemName is designed around one goal: keep a wide expert-parallel EP instance valid across partial failures.

\subsection{Failure Model and Scope}

\SystemName targets fail-stop failures in which a rank becomes permanently or temporarily unreachable due to GPU-process crashes, host failures, or communication link failures. Surviving ranks remain healthy and continue serving. A failed rank may later be repaired or relaunched and should be able to rejoin the instance to restore throughput.

The system assumes that failures are detectable through communication timeouts and that the majority of ranks remain operational. Requests in flight at the moment of failure are reported as failed and must be retried by the client; \SystemName does not buffer or internally retry failed requests, keeping the recovery path simple and the failure semantics explicit. Expert-coverage repair and rank reintegration occur between forward passes, where natural synchronization points already exist in the serving loop.

Out of scope are Byzantine failures, silent data corruption, network partitions that split the cluster into multiple components, and transient performance degradation that does not trigger timeout-based failure detection. These failure modes either require orthogonal detection mechanisms or are better addressed at other layers of the serving stack.

\subsection{Live EP Validity}

The central claim of this paper is that partial-failure tolerance in wide EP should be expressed as a validity condition on a live EP instance rather than as a restart procedure.

\paragraph{Validity contract.}
An EP instance is valid iff three conditions hold simultaneously:
\begin{enumerate}
    \item \textbf{Peer-set validity:} communication targets only currently active and reachable ranks.
    \item \textbf{Expert-coverage validity:} every logical expert is hosted by at least one active rank.
    \item \textbf{Graph-visible routing validity:} the peer and routing state used by CUDA-graph-captured dispatch and combination kernels matches the current active membership and expert placement.
\end{enumerate}

Existing EP stacks bundle these states into one static configuration. \SystemName instead represents them explicitly and repairs them separately. A failure is handled once the instance again satisfies the validity contract, even if it is temporarily operating with reduced capacity.

\subsection{Design Overview}

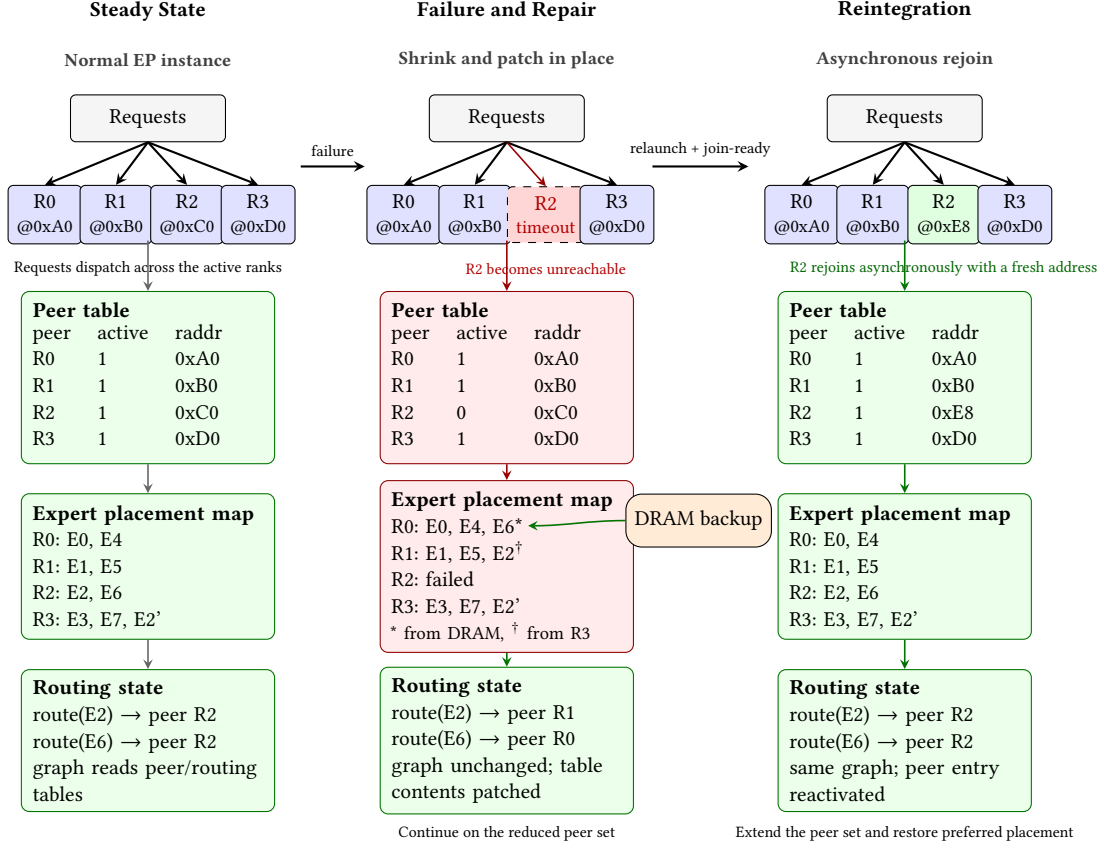
\begin{figure*}
    \centering
    \begin{tikzpicture}[
    x=0.94cm,
    y=0.98cm,
    >=stealth,
    font=\footnotesize,
    title/.style={font=\footnotesize\bfseries},
    sectionlabel/.style={font=\scriptsize\bfseries, text=black!75},
    rank/.style={draw, rounded corners=2pt, minimum width=0.95cm, minimum height=0.62cm, fill=blue!12, align=center},
    rankfailed/.style={draw, dashed, rounded corners=2pt, minimum width=0.95cm, minimum height=0.62cm, fill=red!16, text=red!70!black, align=center},
    rankjoin/.style={draw, rounded corners=2pt, minimum width=0.95cm, minimum height=0.62cm, fill=green!12, align=center},
    req/.style={draw, rounded corners=2pt, minimum width=2.05cm, minimum height=0.62cm, fill=gray!8, align=center},
    dram/.style={draw, rounded corners=6pt, minimum width=1.75cm, minimum height=0.7cm, fill=orange!16, align=center},
    state/.style={draw, rounded corners=3pt, minimum width=3.35cm, minimum height=1.28cm, fill=white, align=left, text width=3.05cm, inner sep=4pt},
    statebad/.style={draw, rounded corners=3pt, minimum width=3.35cm, minimum height=1.28cm, fill=red!8, draw=red!55!black, align=left, text width=3.05cm, inner sep=4pt},
    stategood/.style={draw, rounded corners=3pt, minimum width=3.35cm, minimum height=1.28cm, fill=green!8, draw=green!45!black, align=left, text width=3.05cm, inner sep=4pt},
    note/.style={font=\tiny, align=center},
    smallmono/.style={font=\ttfamily\tiny, align=left},
    flow/.style={->, thick},
    soft/.style={->, semithick, draw=black!60},
    repair/.style={->, semithick, draw=green!45!black},
    danger/.style={->, semithick, draw=red!60!black}
]

\node[title] at (2.3, 8.0) {Steady State};
\node[title] at (7.35, 8.0) {Failure and Repair};
\node[title] at (12.95, 8.0) {Reintegration};

\node[sectionlabel] at (2.3, 7.35) {Normal EP instance};
\node[sectionlabel] at (7.35, 7.35) {Shrink and patch in place};
\node[sectionlabel] at (12.95, 7.35) {Asynchronous rejoin};

\node[req] (req1) at (2.3, 6.55) {Requests};
\node[rank] (s0) at (0.85, 5.25) {R0\\[-1pt]\scriptsize @0xA0};
\node[rank] (s1) at (1.85, 5.25) {R1\\[-1pt]\scriptsize @0xB0};
\node[rank] (s2) at (2.85, 5.25) {R2\\[-1pt]\scriptsize @0xC0};
\node[rank] (s3) at (3.85, 5.25) {R3\\[-1pt]\scriptsize @0xD0};
\draw[flow] (req1.south) -- (s0.north);
\draw[flow] (req1.south) -- (s1.north);
\draw[flow] (req1.south) -- (s2.north);
\draw[flow] (req1.south) -- (s3.north);
\node[note] at (2.3, 4.52) {Requests dispatch across the active ranks};

\node[stategood] (peer1) at (2.3, 3.05) {
    \textbf{Peer table}\\
    \begin{tabular}{@{}lll@{}}
    peer & active & raddr\\
    R0 & 1 & 0xA0\\
    R1 & 1 & 0xB0\\
    R2 & 1 & 0xC0\\
    R3 & 1 & 0xD0
    \end{tabular}
};
\node[stategood] (place1) at (2.3, 0.50) {
    \textbf{Expert placement map}\\
    R0: E0, E4\\
    R1: E1, E5\\
    R2: E2, E6\\
    R3: E3, E7, E2'
};
\node[stategood] (route1) at (2.3, -1.85) {
    \textbf{Routing state}\\
    route(E2) $\rightarrow$ peer R2\\
    route(E6) $\rightarrow$ peer R2\\
    graph reads peer/routing tables
};
\draw[soft] (2.3, 4.9) -- (peer1.north);
\draw[soft] (peer1.south) -- (place1.north);
\draw[soft] (place1.south) -- (route1.north);

\node[req] (req2) at (7.35, 6.55) {Requests};
\node[rank] (f0) at (5.9, 5.25) {R0\\[-1pt]\scriptsize @0xA0};
\node[rank] (f1) at (6.9, 5.25) {R1\\[-1pt]\scriptsize @0xB0};
\node[rankfailed] (f2) at (7.9, 5.25) {R2\\[-1pt]\scriptsize timeout};
\node[rank] (f3) at (8.9, 5.25) {R3\\[-1pt]\scriptsize @0xD0};
\draw[flow] (req2.south) -- (f0.north);
\draw[flow] (req2.south) -- (f1.north);
\draw[flow] (req2.south) -- (f3.north);
\draw[danger] (req2.south) -- (f2.north);
\node[note, text=red!70!black] at (7.9, 4.52) {R2 becomes unreachable};

\node[statebad] (peer2) at (7.35, 3.05) {
    \textbf{Peer table}\\
    \begin{tabular}{@{}lll@{}}
    peer & active & raddr\\
    R0 & 1 & 0xA0\\
    R1 & 1 & 0xB0\\
    R2 & 0 & 0xC0\\
    R3 & 1 & 0xD0
    \end{tabular}
};
\node[statebad] (place2) at (7.35, 0.50) {
    \textbf{Expert placement map}\\
    R0: E0, E4, E6*\\
    R1: E1, E5, E2$^\dagger$\\
    R2: failed\\
    R3: E3, E7, E2'\\
    \scriptsize * from DRAM,\ $^\dagger$ from R3
};
\node[stategood] (route2) at (7.35, -1.85) {
    \textbf{Routing state}\\
    route(E2) $\rightarrow$ peer R1\\
    route(E6) $\rightarrow$ peer R0\\
    graph unchanged; table contents patched
};
\draw[danger] (7.35, 4.9) -- (peer2.north);
\draw[danger] (peer2.south) -- (place2.north);
\draw[repair] (place2.south) -- (route2.north);

\node[dram] (dram2) at (10.05, 1.12) {DRAM backup};
\draw[repair] (dram2.west) to[out=180,in=0] (7.65, 1.05);
\node[note] at (7.35, -3.10) {Continue on the reduced peer set};

\node[req] (req3) at (12.95, 6.55) {Requests};
\node[rank] (j0) at (11.5, 5.25) {R0\\[-1pt]\scriptsize @0xA0};
\node[rank] (j1) at (12.5, 5.25) {R1\\[-1pt]\scriptsize @0xB0};
\node[rankjoin] (j2) at (13.5, 5.25) {R2\\[-1pt]\scriptsize @0xE8};
\node[rank] (j3) at (14.5, 5.25) {R3\\[-1pt]\scriptsize @0xD0};
\draw[flow] (req3.south) -- (j0.north);
\draw[flow] (req3.south) -- (j1.north);
\draw[flow] (req3.south) -- (j2.north);
\draw[flow] (req3.south) -- (j3.north);
\node[note, text=green!40!black] at (13.5, 4.52) {R2 rejoins asynchronously with a fresh address};

\node[stategood] (peer3) at (12.95, 3.05) {
    \textbf{Peer table}\\
    \begin{tabular}{@{}lll@{}}
    peer & active & raddr\\
    R0 & 1 & 0xA0\\
    R1 & 1 & 0xB0\\
    R2 & 1 & 0xE8\\
    R3 & 1 & 0xD0
    \end{tabular}
};
\node[stategood] (place3) at (12.95, 0.50) {
    \textbf{Expert placement map}\\
    R0: E0, E4\\
    R1: E1, E5\\
    R2: E2, E6\\
    R3: E3, E7, E2'
};
\node[stategood] (route3) at (12.95, -1.85) {
    \textbf{Routing state}\\
    route(E2) $\rightarrow$ peer R2\\
    route(E6) $\rightarrow$ peer R2\\
    same graph; peer entry reactivated
};
\draw[repair] (12.95, 4.9) -- (peer3.north);
\draw[repair] (peer3.south) -- (place3.north);
\draw[repair] (place3.south) -- (route3.north);
\node[note] at (12.95, -3.10) {Extend the peer set and restore preferred placement};

\draw[flow] (4.45, 5.9) -- node[above, note] {failure} (5.35, 5.9);
\draw[flow] (9.4, 5.9) -- node[above, note] {relaunch + join-ready} (10.75, 5.9);
\end{tikzpicture}
    \caption{Overview of \SystemName. A rank failure invalidates the active peer set, expert placement, and graph-visible routing state. \SystemName repairs them in place and later reintegrates the recovered rank without rebuilding the EP instance.}
    \label{fig:overview}
\end{figure*}

A rank failure invalidates all three conditions at once: the active peer set changes immediately, some logical experts may disappear with the failed rank, and the graph-visible routing state captured under the old membership no longer describes a correct execution. \SystemName treats recovery as a three-step repair process rather than an instance rebuild.

Figure~\ref{fig:overview} shows the resulting structure. The three repair steps correspond directly to the three tensions introduced in Section~\ref{sec:introduction}. \SystemName restores peer-set validity through a GPU-resident peer table whose entries can be disabled in place while captured graphs continue to execute; restores expert-coverage validity through a repaired placement satisfied by local reuse, GPU-to-GPU relocation, or DRAM-backed reload; and restores graph-visible routing validity through mutable peer and routing tables that match the repaired reduced-capacity configuration. Reintegration is the reverse transition: a deferred-join protocol allows the repaired rank to warm up independently, then the active peer set expands again, the preferred placement is restored, and healthy ranks keep executing the same captured graphs.

Realizing these three repair steps requires coordinated action at two levels. The communication layer (Section~\ref{sec:impl-peer-table} and Section~\ref{sec:impl-reintegration}) owns peer-set validity and graph-visible routing validity through a GPU-resident peer table whose contents can be patched in place and a deferred-join mechanism that keeps recovery asynchronous. The inference system layer (Section~\ref{sec:impl-expert-location} and Section~\ref{sec:impl-backup}) owns expert-coverage validity through explicit expert-location metadata, repaired placement, and weight transfer. Neither layer is self-sufficient: communication repair alone cannot guarantee that every logical expert remains reachable, and inference-layer rerouting alone would still face hanging collectives. The subsections that follow describe both layers as one coordinated recovery mechanism.

\subsection{Membership-Elastic EP Communication}

EP communication sits directly on the critical path of MoE decoding. During each MoE layer, every rank must dispatch token activations to the ranks hosting the selected experts and then collect the results after expert computation. Existing implementations realize this pattern as a collective over a fixed communicator or preconfigured EP group. That design is efficient under a static world, but once one rank disappears the surviving ranks still execute the communication pattern of the old world, so both progress and correctness break. This is Tension~1 in operational form: the communication abstraction assumes a fixed participant set, while the serving runtime needs the effective peer set to change without rebuilding the fast path.

\SystemName restores peer-set validity by expressing EP communication over an addressable device-side peer table rather than a fixed collective group. Each GPU maintains a table of peer contexts in GPU memory. A peer context stores the transport metadata needed to reach that rank and an activity bit indicating whether the rank currently belongs to the active instance. Communication kernels iterate over this table at runtime to issue dispatch, combination, and synchronization operations. Membership is therefore not encoded in communicator identity or in graph structure; it is encoded in mutable device-side state.

This representation has two consequences. First, communication remains GPU-driven. During initialization, \SystemName performs a one-time exchange of the metadata needed to access each peer. Afterwards, CUDA kernels post work queue entries and ring NIC doorbells directly through GPU-mapped RNIC state, so dispatch and combination remain compatible with CUDA-graph execution. Second, membership changes become simple table updates. Failure clears a peer-table entry; reintegration refreshes that entry with new metadata and marks it active again. The straightforward alternatives do not provide this separation: communicator shrink preserves neither graph stability nor application-level EP state, and CPU-managed peer discovery would insert host orchestration back into the common path.

To coordinate progress without reverting to host-side orchestration, \SystemName uses GPU-side synchronization based on RDMA atomics. After issuing its transfers for a communication round, each rank posts atomic updates to its peers and waits until its local GPU-resident counters reflect the expected arrivals from the active peer set. If an expected update does not arrive within a timeout window, the peer is deemed unreachable. Failure detection therefore falls out of the same GPU-side progress mechanism used in steady state, rather than requiring a separate host-managed watchdog in the data path.

\subsection{Repairing Expert Coverage After Failure}
\label{sec:impl-recovery}

Repairing the peer set alone is not enough. In wide EP, a rank failure is also an expert-loss event: some logical experts may become unreachable because their weights were hosted on the failed rank. This is Tension~2 in operational form: the system wants to resume service quickly on surviving ranks, but it cannot safely continue until every logical expert is reachable again. \SystemName therefore treats peer-set repair and expert-coverage repair as separate steps. The former restores communication continuity over the surviving ranks; the latter reconstructs a placement in which every logical expert is again reachable on at least one active rank.

After the communication layer identifies a failed rank, the runtime computes a repaired placement over the surviving ranks. The repair path then satisfies that placement through a hierarchy that follows bandwidth and latency realities. 

\paragraph{Expert weights already reside on the destination rank.}
If the new placement still assigns an expert to a rank that already holds its weights, recovery is purely local. \SystemName updates the local placement metadata and, if needed, reorganizes the weight layout in GPU memory. This is the cheapest recovery path. 

\paragraph{A surviving copy exists on another healthy rank.}
If the destination rank does not already hold the expert but another healthy rank does, \SystemName performs a direct GPU-to-GPU relocation. Because failures often invalidate multiple experts at once, relocation is issued as a batched transfer schedule rather than as a sequence of independent copies. This keeps the repair path aligned with the bandwidth-oriented nature of the underlying fabric.

\paragraph{All GPU-resident copies were lost with the failed rank.}
This is the hard case. Reloading expert weights from persistent storage would dominate recovery time and defeat the purpose of partial-failure tolerance. \SystemName therefore maintains a distributed expert backup service in host DRAM. Each node runs a manager process that stores a subset of expert backups in pinned, RNIC-registered memory. Collectively, these managers hold a full backup copy of the experts needed for recovery. When a surviving rank needs an expert whose live copies were lost, it fetches the backup directly into GPU memory via RDMA, bypassing the filesystem and minimizing CPU involvement.

The hierarchy is deliberate. Local reuse is cheapest, GPU-to-GPU relocation is the common fast path when another replica survives, and DRAM-backed reload is the fallback only when all GPU-resident copies have been lost. By structuring repair as a placement problem with an explicit source hierarchy, \SystemName keeps the common case on fast local or GPU-resident paths and reserves DRAM recovery for the harder cases where redundancy has been exhausted. This resolves Tension~2: degraded-mode recovery becomes fast enough to be useful without sacrificing expert completeness.

\subsection{Rank Reintegration After Repair}
\label{sec:impl-scaling}

Partial-failure tolerance is incomplete if the system can only shrink. In practice, operators want a repaired or relaunched rank to rejoin the existing instance so that throughput can return to its pre-failure level. \SystemName therefore supports rank reintegration: healthy ranks continue serving with reduced capacity while the failed rank is relaunched asynchronously, and the rank is incorporated back into the live instance once it becomes ready. This is Tension~3 in operational form: the recovering rank needs a long local warmup path---reinitializing its Python runtime, CUDA state, communication endpoints, and model weights---but healthy ranks cannot be forced to wait for it.

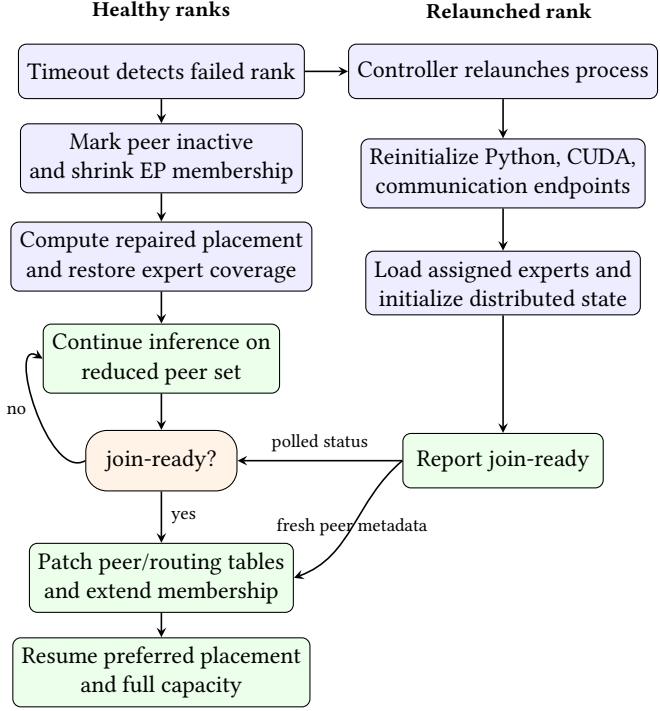
\begin{figure}
    \centering
    \begin{tikzpicture}[
    >=stealth,
    font=\small,
    lane/.style={font=\footnotesize\bfseries},
    flowbox/.style={draw, rounded corners=3pt, minimum width=2.65cm, minimum height=0.72cm, align=center, fill=blue!7},
    flowboxgood/.style={draw, rounded corners=3pt, minimum width=2.65cm, minimum height=0.72cm, align=center, fill=green!8},
    judge/.style={draw, rounded corners=8pt, minimum width=2.0cm, minimum height=0.8cm, align=center, fill=orange!10},
    arrow/.style={->, semithick},
    side/.style={font=\scriptsize, align=center}
]

\node[lane] at (0, 5.7) {Healthy ranks};
\node[lane] at (4.6, 5.7) {Relaunched rank};

\node[flowbox] (detect) at (0, 4.9) {Timeout detects failed rank};
\node[flowbox] (shrink) at (0, 3.75) {Mark peer inactive\\and shrink EP membership};
\node[flowbox] (repair) at (0, 2.45) {Compute repaired placement\\and restore expert coverage};
\node[flowboxgood] (serve) at (0, 1.1) {Continue inference on\\reduced peer set};
\node[judge] (ready) at (0, -0.25) {join-ready?};
\node[flowboxgood] (patch) at (0, -1.80) {Patch peer/routing tables\\and extend membership};
\node[flowboxgood] (restore) at (0, -3.05) {Resume preferred placement\\and full capacity};

\node[flowbox] (relaunch) at (4.52, 4.9) {Controller relaunches process};
\node[flowbox] (setup) at (4.52, 3.55) {Reinitialize Python, CUDA,\\communication endpoints};
\node[flowbox] (state) at (4.52, 2.1) {Load assigned experts and\\initialize distributed state};
\node[flowboxgood] (signal) at (4.52, -0.25) {Report join-ready};

\draw[arrow] (detect) -- (shrink);
\draw[arrow] (shrink) -- (repair);
\draw[arrow] (repair) -- (serve);
\draw[arrow] (serve) -- (ready);
\draw[arrow] (ready) -- node[right, side] {yes} (patch);
\draw[arrow] (patch) -- (restore);
\draw[arrow] (relaunch) -- (setup);
\draw[arrow] (setup) -- (state);
\draw[arrow] (state) -- (signal);

\draw[arrow] (detect.east) -- (relaunch.west);
\draw[arrow] (signal.west) -- node[above, side] {polled status} (ready.east);
\draw[arrow] (ready.west) to[out=210,in=150] node[left, side] {no} (serve.west);
\draw[arrow] (signal.west) to[out=220,in=20] node[side] {fresh peer metadata} (patch.east);
\end{tikzpicture}
    \caption{Failure and reintegration workflow in \SystemName. Healthy ranks restore a reduced-capacity instance first, while the failed rank reinitializes independently and later rejoins through deferred join.}
    \label{fig:reintegration-seq}
\end{figure}

Reintegration is the reverse of failure handling, but the main implementation challenge is not merely refreshing communication metadata. It is allowing the recovering process to execute its full warmup path asynchronously while the rest of the instance keeps serving. A controller restarts the failed process outside the serving critical path. The restarted rank reinitializes its runtime, CUDA state, communication endpoints, and the expert weights assigned to it under the restored placement. Meanwhile, the healthy ranks continue inference on the reduced peer set and periodically poll whether the relaunched rank has reached a join-ready state. The rightmost column of Figure~\ref{fig:overview} depicts this asynchronous rejoin path, while Figure~\ref{fig:reintegration-seq} makes the ordering explicit. The rejected alternative is the standard distributed-runtime path in which the recovering rank rebuilds process-group state and recaptures graphs with the rest of the group participating. That path would turn reintegration into another global pause.

The main technical challenge is that reintegration changes EP membership again, but the running instance should not pay the cost of CUDA-graph recapture on healthy ranks. The hard part is therefore to extend the distributed-runtime semantics so the recovering rank can initialize alone and join only after it is locally ready, as described in Section~\ref{sec:impl-reintegration}. \SystemName provides this through a deferred-join backend path: the recovering rank forms a local-only process group, rebuilds its CUDA and communication state, loads weights, and recaptures its graphs in isolation. Only after that warmup completes do the healthy ranks incorporate it by refreshing peer metadata, extending the active-membership mask, and updating routing-visible state between forward passes.

This property is the key to the reintegration story. The expensive warmup work is not eliminated; it is moved off the healthy-rank critical path. A partial failure first contracts the active peer set and restores validity on the reduced instance. Later, a repaired rank expands the peer set again. Both transitions operate over the same abstractions---an explicit peer set, an explicit expert-placement map, and graph-stable peer and routing tables---rather than over a fixed-world communicator. This resolves Tension~3: reintegration becomes a bounded join event on healthy ranks rather than a second rebuild event.

\section{Membership-Elastic EP Communication}
\label{sec:implementation}

This section and Section~\ref{sec:sglanglayer} describe how the design in
Section~\ref{sec:design} is realized as a two-layer serving stack. This section covers the                        
communication-library layer: membership-elastic EP communication implemented in
Mooncake, specifically Mooncake EP and Mooncake PG.\footnote{\url{https://github.com/kvcache-ai/Mooncake}}
Section~\ref{sec:sglanglayer} then covers the SGLang layer,\footnote{\url{https://github.com/sgl-project/sglang}} which
builds inference-layer repair on top of this elastic communication substrate.
The implementation spans roughly 10\,K lines across Mooncake and SGLang. We
describe the communication layer first because it is what makes inference-layer
repair possible without halting the serving critical path.

\SystemName's communication layer must satisfy two properties simultaneously:
EP dispatch and combination must remain GPU-driven and CUDA-graph-compatible even
as the active peer set changes, and a recovering rank must rebuild its
communication state in isolation without pulling healthy ranks into barriers or
process-group setup. The GPU-resident peer table
(Section~\ref{sec:impl-peer-table}) satisfies the first requirement; the
deferred-join mechanism (Section~\ref{sec:impl-reintegration}) satisfies the
second. Together they close Tensions~1 and~3.

\subsection{GPU-Resident Peer Table and Connection Management}
\label{sec:impl-peer-table}

\begin{figure}[t]
\centering
\begin{minipage}{\linewidth}
{\tt \small
\begin{verbatim}
// Per-rank entry; lives entirely in GPU vRAM
struct peer_entry {
    int32_t  active;    // 0 = failed, skip this peer
    int32_t  nvlink;    // 1 = NVLink-reachable
    void    *ipc_ptr;   // IPC-mapped buffer (NVLink path)
    qp_devctx_t qp;     // IBGDA context  (RDMA path)
};

struct qp_devctx {
    uint32_t qpn;
    void    *wq, *cq;   // work/completion queues
    void    *dbr;       // doorbell register (GPU-mapped)
    uint16_t wq_head, wq_tail;
};
\end{verbatim}
}
\end{minipage}
\vspace{2pt}\hrule\vspace{4pt}
\begin{minipage}{\linewidth}
{\tt \small
\begin{verbatim}
// Dispatch kernel (simplified): runs entirely on GPU
for each (token t, dst_rank r):
    e = peer_table[r]
    if e.active == 0: continue     // skip failed peer
    if e.nvlink:
        write t to e.ipc_ptr       // NVLink fast path
    else:
        post_wqe(e.qp, t)
        ring_doorbell(e.qp.dbr)    // no CPU involvement
    // signal sent; receiver polls its own counter;
    // timeout -> set e.active = 0 and continue
\end{verbatim}
}
\end{minipage}
\caption{\SystemName's GPU-resident peer-table entry (top) and simplified
    dispatch path (bottom). Membership changes update table contents rather
    than graph structure, so the same captured kernels survive failure and
    reintegration.}
\label{fig:peer-table}
\end{figure}

The device-side peer table is \SystemName's central data structure for EP communication. Each EP buffer allocates a contiguous GPU-memory array indexed by rank; each entry stores the remote buffer address, memory key, IBGDA queue-pair context, NVLink reachability flag, and an IPC-exported pointer for intra-node access. CUDA kernels index this array directly at dispatch and combination time, so every routing decision stays in device memory rather than in host-side lookup code or graph structure.

That organization is what keeps the path high-performance but still elastic: the dispatch kernel can load one rank-indexed entry, choose NVLink or GPU-initiated RDMA, and skip failed peers by testing a single \texttt{active} bit. The same kernel binary therefore handles steady state, degraded execution, and restored configuration; membership changes update table contents, not captured kernel structure.

Connection setup runs once during initialization. \SystemName all-gathers memory-region metadata and queue-pair identifiers, programs the IBGDA endpoints, exchanges IPC handles, and fills the GPU-resident table. Dispatch and combination then follow the design in prior work~\cite{deepep}: dispatch writes payloads through NVLink or GPU-initiated RDMA depending on the entry, combination waits for signals from active peers and reduces expert outputs, and double-buffered staging regions overlap communication with compute.

\paragraph{Failure detection from the GPU side.}
Progress within a communication round is tracked through per-rank signal counters in GPU memory. After writing its payload, the sender increments the receiver's counter via an RDMA atomic or a release-store on NVLink paths. The receiver polls its own counter from within the CUDA kernel. If the expected increment does not arrive within a timeout (currently 1\,s), the kernel clears the \texttt{active} bit for that peer and continues without it. The failure report propagates to the host only after the kernel exits, so failure detection requires no separate heartbeat or host-driven watchdog. This design assumes fail-stop failures; transient slowdowns that exceed the timeout threshold are treated as failures.

\paragraph{In-place peer-table updates for membership changes.}
When membership changes---either a rank failure or a rank reintegration---\SystemName updates the GPU-resident table in place rather than reallocating the EP buffer. For a failing rank, the \texttt{active} bit is cleared. For a rejoining rank, \SystemName re-exchanges RDMA metadata for that rank, reprograms its queue-pair entry, and sets the bit. The dispatch and combine kernels are unchanged; only the table they read changes.

\subsection{Rank Reintegration Without Graph Recapture}
\label{sec:impl-reintegration}

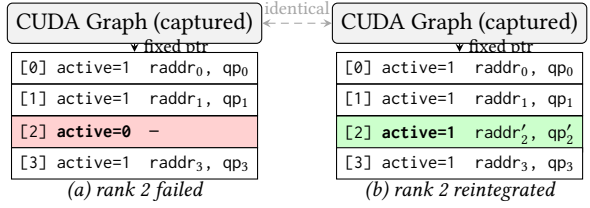
\begin{figure}[t]
\centering
\begin{tikzpicture}[>=stealth, font=\small,
  cg/.style={draw, rounded corners=2pt, fill=gray!10,
             minimum width=3.1cm, minimum height=0.52cm, align=center},
  pe/.style={draw, minimum width=3.1cm, minimum height=0.40cm,
             inner sep=3pt, font=\ttfamily\scriptsize, align=left}
]

\node[cg] (G1) at (0,0) {CUDA Graph (captured)};
\draw[->] (G1.south) -- node[right, font=\scriptsize] {fixed ptr} (0,-0.38);

\node[pe, fill=white,  anchor=north] at (0,-0.38) {[0]\ active=1\ \ raddr$_0$, qp$_0$};
\node[pe, fill=white,  anchor=north] at (0,-0.80) {[1]\ active=1\ \ raddr$_1$, qp$_1$};
\node[pe, fill=red!18, anchor=north] at (0,-1.22) {[2]\ \textbf{active=0}\ \ ---\phantom{addr$_2$, qp$_2$}};
\node[pe, fill=white,  anchor=north] at (0,-1.64) {[3]\ active=1\ \ raddr$_3$, qp$_3$};

\node[font=\footnotesize\itshape] at (0,-2.22) {(a) rank 2 failed};

\node[cg] (G2) at (4.3,0) {CUDA Graph (captured)};
\draw[->] (G2.south) -- node[right, font=\scriptsize] {fixed ptr} (4.3,-0.38);

\node[pe, fill=white,    anchor=north] at (4.3,-0.38) {[0]\ active=1\ \ raddr$_0$, qp$_0$};
\node[pe, fill=white,    anchor=north] at (4.3,-0.80) {[1]\ active=1\ \ raddr$_1$, qp$_1$};
\node[pe, fill=green!20, anchor=north] at (4.3,-1.22) {[2]\ \textbf{active=1}\ \ raddr$_2'$, qp$_2'$};
\node[pe, fill=white,    anchor=north] at (4.3,-1.64) {[3]\ active=1\ \ raddr$_3$, qp$_3$};

\node[font=\footnotesize\itshape] at (4.3,-2.22) {(b) rank 2 reintegrated};

\draw[<->, densely dashed, gray!70]
  (G1.east) -- node[above, font=\scriptsize] {identical} (G2.west);

\end{tikzpicture}
\caption{Peer-table update during rank reintegration. The graph keeps a fixed
    pointer to the table; only the rejoining rank's entry is refreshed.}
\label{fig:reintegration-peer-table}
\end{figure}

When a failed rank is relaunched, it reinitializes its CUDA state, rebuilds its communication endpoints, loads its expert weights, and signals readiness to the cluster. The controller runs this process outside the serving critical path, while healthy ranks continue inference on the reduced peer set.

\paragraph{Deferred process-group join for async recovery.}
The recovered rank's initialization is fully asynchronous because \SystemName provides a custom \texttt{torch.distributed} backend that supports elastic membership. This backend allows a recovering rank to construct a local-only process group with only itself marked active, satisfying upper layer system's distributed-runtime requirements without connectivity to healthy ranks. Without this deferred-join capability, weight loading, metadata broadcasts, and synchronization barriers would require healthy ranks to participate and would pause inference. Instead, the recovered rank loads model weights and recaptures CUDA graphs in isolation. Only when local initialization completes does it call an explicit join operation, which publishes metadata to the peers, registers with the connection poller, and waits for full-group connectivity.

\paragraph{Healthy-side reintegration steps.}
Reintegration from the healthy side consists of two steps. First, \SystemName re-initializes the RDMA connection to the recovered rank by re-exchanging metadata, programming the IBGDA endpoint, and updating that rank's peer-table entry. Second, the current expert-location metadata is broadcasted to the recovered rank to ensure all ranks share the same view of expert-location allocation.

The key property is that neither step rebuilds any CUDA graph on the healthy ranks. The graphs hold a fixed device pointer to the peer table, and reintegration only patches table contents between forward passes. The only visible effect to inference is that the rank appears in the active bitmap and its updated entry becomes live for the next dispatch round.


\section{Expert-Coverage Repair}
\label{sec:sglanglayer}

The communication layer restores peer-set validity and graph-visible routing
validity after a failure, but not expert-coverage validity: a rank failure is
also an expert-loss event. This section describes the two inference-system-layer
mechanisms that close Tension~2: explicit expert-location metadata that makes
coverage a first-class runtime state, and a DRAM-backed expert backup service
that bounds weight-recovery cost when all GPU-resident copies of an expert have
been lost. Both are implemented in our Mooncake+SGLang serving stack, but
the abstractions themselves are not specific to SGLang.

\subsection{Expert-Location Metadata and Placement Repair}
\label{sec:impl-expert-location}

\SystemName maintains explicit expert-location metadata that maps every physical expert slot to a logical expert and records all physical locations where each logical expert resides. This metadata is the shared source of truth for the dispatcher, load balancer, and repair path. Keeping it explicit---rather than deriving it from communicator structure---is what allows expert coverage to be reasoned about independently of which ranks are alive.

The placement-repair path runs when the active-rank bitmap changes. \SystemName invokes an elasticity-aware variant of the expert-parallelism load balancer (EPLB), which takes the current active-rank set as input and returns a placement that covers all logical experts over the surviving ranks. The repair path then moves weights to match that placement, following the three-case hierarchy from Section~\ref{sec:impl-recovery}.

The active-rank bitmap is consulted atomically before each transfer is issued: if a needed source rank has gone inactive since the schedule was built, the corresponding expert is recorded as missing immediately, and the DRAM-backup path fills it in.

\subsection{Distributed DRAM-Backed Expert Backups}
\label{sec:impl-backup}

The backup service provides the fallback path when all GPU-resident copies of an expert are lost. Rather than reading from a checkpoint on a distributed filesystem, each node keeps a pre-loaded, RNIC-registered copy of a subset of expert weights in pinned CPU DRAM. The service is optional and trades CPU DRAM capacity for faster recovery; for the 671\,B-parameter model used in our evaluation, each of the four nodes holds about 167.5\,GB of model weights in DRAM.

Each node runs a standalone backup manager process that loads its assigned experts at startup, assembles them into a contiguous pinned buffer, registers the buffer for RDMA, and publishes a descriptor table mapping expert identifiers to addresses and sizes. When the repair path reports missing experts, a backup client consults that table and issues batched GPU-initiated RDMA reads directly into GPU parameter tensors.


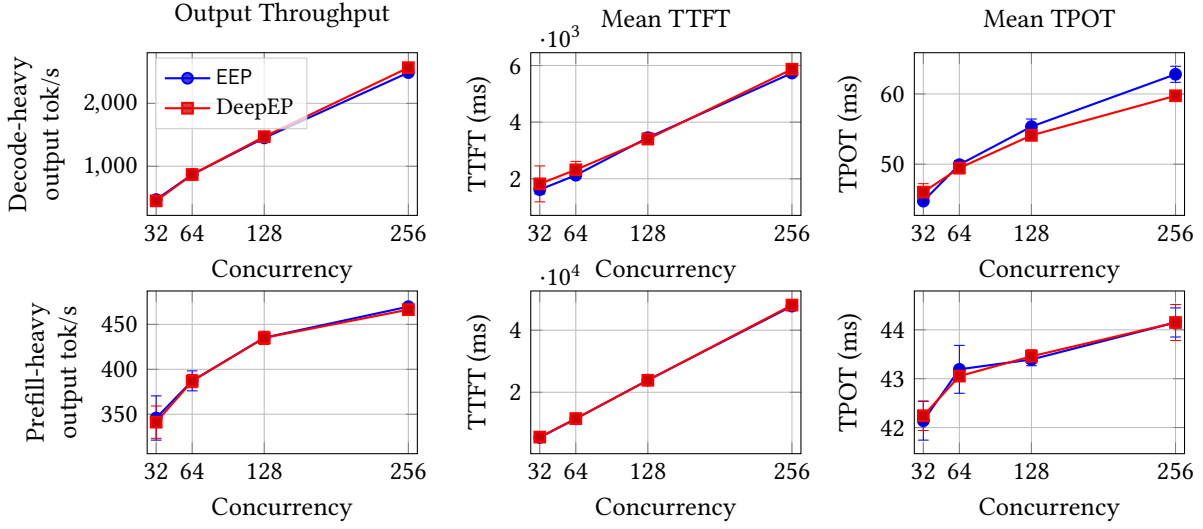
\begin{figure*}[t]
    \centering
    \begin{tikzpicture}
        \begin{groupplot}[
            group style={
                group size=3 by 2,
                horizontal sep=1.5cm,
                vertical sep=1.0cm,
            },
            width=0.29\textwidth,
            height=0.21\textwidth,
            xmin=24, xmax=264,
            xtick={32,64,128,256},
            xlabel={Concurrency},
            grid=major,
            error bars/y dir=both,
            error bars/y explicit,
            legend style={
                font=\small,
                cells={anchor=west},
                draw=none,
            },
            every axis plot/.append style={
                thick,
                mark options={solid},
            },
        ]
            \nextgroupplot[
                title={Output Throughput},
                ylabel={\shortstack{Decode-heavy\\output tok/s}},
                legend style={
                    at={(0.03,0.97)},
                    anchor=north west,
                    draw=black!20,
                    fill=white,
                    fill opacity=0.75,
                    text opacity=1,
                },
            ]
                \addplot+[
                    color=blue,
                    mark=*,
                ] table[
                    x=concurrency,
                    y=mooncake_output_throughput_mean,
                    y error=mooncake_output_throughput_std,
                    col sep=comma,
                ] {experiments/static_serving_performance/plots/data/decodeheavy.csv};
                \addplot+[
                    color=red,
                    mark=square*,
                ] table[
                    x=concurrency,
                    y=deepep_output_throughput_mean,
                    y error=deepep_output_throughput_std,
                    col sep=comma,
                ] {experiments/static_serving_performance/plots/data/decodeheavy.csv};
                \legend{\SystemName,DeepEP}

            \nextgroupplot[
                title={Mean TTFT},
                ylabel={\shortstack{TTFT (ms)}},
                ytick={2000,4000,6000},
                scaled y ticks=base 10:-3,
            ]
                \addplot+[
                    color=blue,
                    mark=*,
                ] table[
                    x=concurrency,
                    y=mooncake_mean_ttft_ms_mean,
                    y error=mooncake_mean_ttft_ms_std,
                    col sep=comma,
                ] {experiments/static_serving_performance/plots/data/decodeheavy.csv};
                \addplot+[
                    color=red,
                    mark=square*,
                ] table[
                    x=concurrency,
                    y=deepep_mean_ttft_ms_mean,
                    y error=deepep_mean_ttft_ms_std,
                    col sep=comma,
                ] {experiments/static_serving_performance/plots/data/decodeheavy.csv};

            \nextgroupplot[
                title={Mean TPOT},
                ylabel={\shortstack{TPOT (ms)}},
            ]
                \addplot+[
                    color=blue,
                    mark=*,
                ] table[
                    x=concurrency,
                    y=mooncake_mean_tpot_ms_mean,
                    y error=mooncake_mean_tpot_ms_std,
                    col sep=comma,
                ] {experiments/static_serving_performance/plots/data/decodeheavy.csv};
                \addplot+[
                    color=red,
                    mark=square*,
                ] table[
                    x=concurrency,
                    y=deepep_mean_tpot_ms_mean,
                    y error=deepep_mean_tpot_ms_std,
                    col sep=comma,
                ] {experiments/static_serving_performance/plots/data/decodeheavy.csv};

            \nextgroupplot[
                ylabel={\shortstack{Prefill-heavy\\output tok/s}},
            ]
                \addplot+[
                    color=blue,
                    mark=*,
                ] table[
                    x=concurrency,
                    y=mooncake_output_throughput_mean,
                    y error=mooncake_output_throughput_std,
                    col sep=comma,
                ] {experiments/static_serving_performance/plots/data/prefillheavy.csv};
                \addplot+[
                    color=red,
                    mark=square*,
                ] table[
                    x=concurrency,
                    y=deepep_output_throughput_mean,
                    y error=deepep_output_throughput_std,
                    col sep=comma,
                ] {experiments/static_serving_performance/plots/data/prefillheavy.csv};

            \nextgroupplot[
                ylabel={\shortstack{TTFT (ms)}},
            ]
                \addplot+[
                    color=blue,
                    mark=*,
                ] table[
                    x=concurrency,
                    y=mooncake_mean_ttft_ms_mean,
                    y error=mooncake_mean_ttft_ms_std,
                    col sep=comma,
                ] {experiments/static_serving_performance/plots/data/prefillheavy.csv};
                \addplot+[
                    color=red,
                    mark=square*,
                ] table[
                    x=concurrency,
                    y=deepep_mean_ttft_ms_mean,
                    y error=deepep_mean_ttft_ms_std,
                    col sep=comma,
                ] {experiments/static_serving_performance/plots/data/prefillheavy.csv};

            \nextgroupplot[
                ylabel={\shortstack{TPOT (ms)}},
            ]
                \addplot+[
                    color=blue,
                    mark=*,
                ] table[
                    x=concurrency,
                    y=mooncake_mean_tpot_ms_mean,
                    y error=mooncake_mean_tpot_ms_std,
                    col sep=comma,
                ] {experiments/static_serving_performance/plots/data/prefillheavy.csv};
                \addplot+[
                    color=red,
                    mark=square*,
                ] table[
                    x=concurrency,
                    y=deepep_mean_tpot_ms_mean,
                    y error=deepep_mean_tpot_ms_std,
                    col sep=comma,
                ] {experiments/static_serving_performance/plots/data/prefillheavy.csv};
        \end{groupplot}
    \end{tikzpicture}
    \caption{Static serving comparison between \SystemName and a fixed-membership DeepEP baseline in the same PD-disaggregated deployment. Error bars show standard deviation over three repetitions.}
    \label{fig:perf-static}
\end{figure*}

\section{Evaluation}

We evaluate \SystemName to answer three questions that map directly to the live EP validity contract introduced in Section~\ref{sec:design}:

\begin{enumerate}
    \item Does explicit mutable membership introduce overhead under static serving? (\S~\ref{sec:perf-static})
    \item How quickly can \SystemName restore a valid reduced-capacity instance after a partial failure? (\S~\ref{sec:perf-fault})
    \item Can a repaired rank reintegrate without forcing healthy ranks through instance rebuild? (\S~\ref{sec:perf-reintegration})
\end{enumerate}

\subsection{Experimental Setup}

\paragraph{Testbed}

All experiments are conducted on a cluster of 6 compute nodes, each equipped with 8 $\times$ NVIDIA GPUs interconnected by NVLink within the node and 8 $\times$ 400 Gbps Mellanox RoCE links across nodes. Each node runs Ubuntu 24.04.2 LTS, CUDA 12.9, driver 535.161.08, and Mellanox OFED 23.10 with GPUDirect RDMA enabled.

\paragraph{Implementation}

We implemented \SystemName with PyTorch 2.8.0 and Python 3.12, comprising roughly 10\,K lines in total: approximately 7.6\,K lines of C++, CUDA, and Python covering the membership-elastic communication substrate, the expert-coverage repair path, and the reintegration control path; and approximately 2.3\,K lines of Python in SGLang covering elastic-EP runtime integration, expert-location management, and the Mooncake EP dispatcher. We preserved API compatibility with existing MoE serving workflows.

\paragraph{Model and Workload}

We evaluate with DeepSeek-V3~\cite{deepseekai2024deepseekv3technicalreport} (built upon the DeepSeekMoE architecture~\cite{deepseekmoe} and DeepSeek-V2~\cite{deepseekv2}), a 671B-parameter MoE model with 256 experts. The static-performance study uses decode-heavy (1024-in / 1024-out) and prefill-heavy (4096-in / 256-out) workloads at concurrencies from 32 to 256. The failure-recovery and reintegration studies use a decode-heavy configuration of 256 input tokens and 4096 output tokens in closed-loop at maximum concurrency 512.

\paragraph{Baselines}

For steady-state performance, we compare against DeepEP, a fixed-membership EP backend, under the same PD-disaggregated SGLang deployment. For failure scenarios, the fixed-membership backend baseline does not support partial-failure recovery; we measure the full-instance restart time on the same hardware, which is the only available recovery path.

\paragraph{Failure Injection}

We simulate rank failures by terminating GPU processes via \texttt{SIGKILL}. Post-recovery throughput is read from the throughput time series after the system returns to reduced-capacity steady-state service.

\begin{figure*}[t]
    \centering
    \input{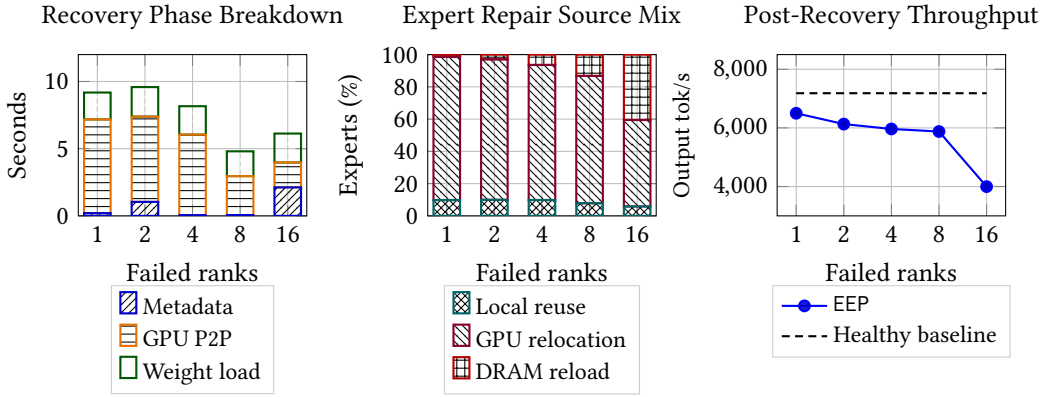}
    \begin{tikzpicture}
        \begin{groupplot}[
            group style={
                group size=3 by 1,
                horizontal sep=1.6cm,
            },
            width=0.17\textwidth,
            height=0.12\textwidth,
            scale only axis,
            symbolic x coords={1,2,4,8,16},
            xtick=data,
            xlabel={Failed ranks},
            grid=major,
            xlabel style={yshift=-1pt},
            legend style={
                font=\small,
                cells={anchor=west},
                draw=none,
            },
        ]
            \nextgroupplot[
                title={Recovery Phase Breakdown},
                ylabel={Seconds},
                ymin=0,
                ymax=12,
                ybar stacked,
                bar width=10pt,
                every axis plot/.append style={thick},
                legend style={
                    at={(0.5,-0.43)},
                    anchor=north,
                    legend columns=1,
                    draw=black!20,
                    fill=white,
                    fill opacity=0.9,
                    text opacity=1,
                },
            ]
                \addplot+[fill=blue!65, fill opacity=0.95, draw=blue!80!black, pattern=north east lines, pattern color=black, mark=none] table[
                    x=failed_ranks,
                    y=metadata_exchange_sec,
                    col sep=comma,
                ] {experiments/failure_recovery_performance/plots/data/failure_recovery.csv};
                \addplot+[fill=orange!85, fill opacity=0.95, draw=orange!90!black, pattern=horizontal lines, pattern color=black, mark=none] table[
                    x=failed_ranks,
                    y=gpu_p2p_exchange_sec,
                    col sep=comma,
                ] {experiments/failure_recovery_performance/plots/data/failure_recovery.csv};
                \addplot+[fill=green!60!black, fill opacity=0.95, draw=green!35!black, pattern=dots, pattern color=white, mark=none] table[
                    x=failed_ranks,
                    y=weight_loading_sec,
                    col sep=comma,
                ] {experiments/failure_recovery_performance/plots/data/failure_recovery.csv};
                \legend{Metadata,GPU P2P,Weight load}

            \nextgroupplot[
                title={Expert Repair Source Mix},
                ylabel={Experts (\%)},
                ymin=0,
                ymax=100,
                ybar stacked,
                bar width=10pt,
                every axis plot/.append style={thick},
                legend style={
                    at={(0.5,-0.43)},
                    anchor=north,
                    legend columns=1,
                    draw=black!20,
                    fill=white,
                    fill opacity=0.9,
                    text opacity=1,
                },
            ]
                \addplot+[fill=teal!65, fill opacity=0.95, draw=teal!80!black, pattern=crosshatch, pattern color=black, mark=none] table[
                    x=failed_ranks,
                    y=local_reuse_pct,
                    col sep=comma,
                ] {experiments/failure_recovery_performance/plots/data/failure_recovery.csv};
                \addplot+[fill=purple!55, fill opacity=0.95, draw=purple!70!black, pattern=north west lines, pattern color=black, mark=none] table[
                    x=failed_ranks,
                    y=gpu_relocation_pct,
                    col sep=comma,
                ] {experiments/failure_recovery_performance/plots/data/failure_recovery.csv};
                \addplot+[fill=red!65, fill opacity=0.95, draw=red!70!black, pattern=grid, pattern color=black, mark=none] table[
                    x=failed_ranks,
                    y=dram_reload_pct,
                    col sep=comma,
                ] {experiments/failure_recovery_performance/plots/data/failure_recovery.csv};
                \legend{Local reuse,GPU relocation,DRAM reload}

            \nextgroupplot[
                title={Post-Recovery Throughput},
                ylabel={Output tok/s},
                ymin=3000,
                ymax=8500,
                every axis plot/.append style={thick, mark options={solid}},
                ylabel style={font=\small},
                legend style={
                    at={(0.5,-0.43)},
                    anchor=north,
                    legend columns=1,
                    draw=black!20,
                    fill=white,
                    fill opacity=0.9,
                    text opacity=1,
                },
            ]
                \addplot+[color=blue, mark=*] table[
                    x=failed_ranks,
                    y=post_output_throughput,
                    col sep=comma,
                    unbounded coords=jump,
                ] {experiments/failure_recovery_performance/plots/data/failure_recovery.csv};
                \addplot+[color=black, densely dashed, mark=none] coordinates {
                    (1,\FailureRecoveryHealthyThroughput)
                    (16,\FailureRecoveryHealthyThroughput)
                };
                \legend{\SystemName, Healthy baseline}
        \end{groupplot}
    \end{tikzpicture}
    \caption{Failure-recovery results under the 2-prefill, 4-decode deployment with 256-input, 4096-output requests and maximum concurrency 512. Left: wall-clock maxima of the three recovery phases. Middle: aggregated expert repair-source mix across surviving ranks; GPU relocation dominates through the eight-rank case, with the DRAM-reload share rising as replica scarcity grows. Right: post-recovery throughput measured from the reduced-capacity serving interval in the throughput time series after the recovery window closes; the dashed line shows the pre-fault healthy throughput.}
    \label{fig:recovery-time}
\end{figure*}

\begin{figure*}[t]
    \centering
    \includegraphics[width=0.8\textwidth]{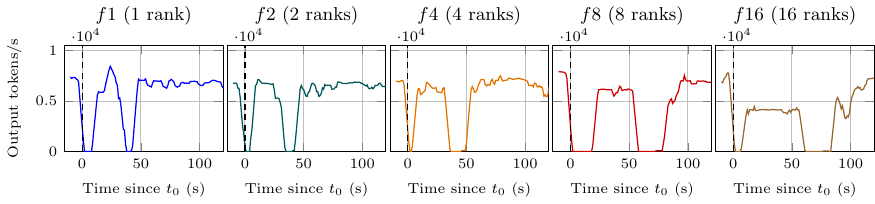}
    \caption{Rank-reintegration throughput traces. Each panel shows one failure scale ($f1$–$f16$); the dashed vertical line marks $t_0$, the first zero-throughput second after the run reaches steady state. Traces use a 5-second moving average. The second zero-throughput windows are 4\,s ($f1$), 6\,s ($f2$), 9\,s ($f4$), 15\,s ($f8$), and 15\,s ($f16$).}
    \label{fig:reintegration-trace}
\end{figure*}

\subsection{Static Serving Performance}
\label{sec:perf-static}

The steady-state question is whether the additional indirection introduced by \SystemName---routing each EP dispatch and combination through a GPU-resident peer table rather than a fixed communicator---materially perturbs throughput or latency relative to a backend that encodes membership as static structure.

Figure~\ref{fig:perf-static} shows that it does not. Across the full concurrency sweep, \SystemName stays within 4.4\% of DeepEP throughput overall and within about 1.4\% on the prefill-heavy workload; the direction of the difference varies across points and is consistent with run-to-run noise rather than a systematic penalty. TTFT and TPOT also track closely at mid-to-high concurrencies.

The dominant cost of EP serving is token dispatch and combination across GPUs, and the extra table lookup per round is negligible relative to that cost. The steady-state path through \SystemName is otherwise structurally the same as through a fixed-membership backend.

\subsection{Failure Recovery Performance}
\label{sec:perf-fault}

The failure-recovery question is whether \SystemName can restore a semantically valid instance quickly enough to matter. The comparison is full-instance restart, the only recovery path available in a fixed-membership design, which on this workload takes 348\,s.

\paragraph{Recovery is fast and bounded.}
Figure~\ref{fig:recovery-time} shows recovery times across failure scales ranging from one failed rank to sixteen, all under the same 2-prefill, 4-decode deployment. Recovery completes in roughly 6--11\,s across the tested range. The exact totals vary somewhat across failure scales, but all cases remain far below the 348\,s full-instance restart baseline. The improvement over full-instance restart is 1 to 2 orders of magnitude.

\paragraph{The repair-source hierarchy behaves as designed.}
Figure~\ref{fig:recovery-time} (middle) shows how the mix of repair sources shifts with failure scale. At small scales, most expert weights are relocated directly between GPUs on surviving ranks, so recovery is dominated by GPU-to-GPU transfer time. As failure scale grows and surviving copies become scarcer, a larger fraction of experts must be reloaded from the distributed DRAM backup service. This shift explains the overall phase balance in the recovery breakdown.

\paragraph{Reduced-capacity service remains productive.}
Figure~\ref{fig:recovery-time} (right) shows post-recovery throughput relative to the pre-fault baseline of 7.2\,K output tokens/s. The recovered instance sustains a meaningful fraction of healthy throughput across all failure scales and continues serving while failed ranks are repaired asynchronously.

\subsection{Rank Reintegration Performance}
\label{sec:perf-reintegration}

Figure~\ref{fig:intro-reintegration-hero} in the introduction already shows the single-rank failure case end-to-end: two bounded pauses---one for recovery and one for reintegration---versus a 348\,s full restart. This section asks whether the same structure holds across a range of failure scales, and whether it does so without pulling healthy ranks off the serving critical path.

\paragraph{Both interruption windows are bounded across all tested scales.}
Figure~\ref{fig:reintegration-trace} shows throughput time series for each failure scale, aligned to the moment of the failure event. Each trace exhibits the same structure: a pre-fault steady-state region, a bounded zero-throughput window as the recovery procedure runs, a reduced-capacity serving plateau, and a second bounded window as the repaired rank reintegrates. Both interruption windows remain in the single-digit-to-low-tens-of-seconds range at all tested scales. Neither event triggers a rebuild of the healthy ranks' execution state.

\paragraph{Reintegration does not rebuild the healthy ranks.}
The second zero-throughput window is the direct test of the design claim for Tension~3. When a repaired rank is ready to rejoin, healthy ranks patch the GPU-resident peer and routing tables in place and extend the active-membership mask---no CUDA-graph recapture, no global barrier, and no model-state reload. The recovering rank performs its full reinitialization in isolation, off the healthy-rank critical path. The bounded second pause shows that healthy ranks incorporate the rejoining rank without reentering instance initialization.

\paragraph{The contrast with full-instance restart.}
The full-restart baseline, which requires 348\,s on the same workload, corresponds to a single uninterrupted outage. \SystemName's traces instead show two brief interruptions separated by a functional serving plateau. The total time off-service is far shorter, and the interval between the pauses represents real productive throughput rather than forced downtime.
\section{Related Work}

Prior work targets static EP execution, planned elasticity, or communicator-level recovery; \SystemName instead targets live partial-failure repair for wide-EP serving without rebuilding the service.

\paragraph{LLM Serving Systems}

Modern LLM serving systems such as vLLM and SGLang improve throughput and latency through batching, KV-cache management, scheduling, and efficient execution \cite{vllm,sglang}. Both support MoE serving, and vLLM additionally exposes EP backends, expert load balancing \cite{eplb}, and an experimental elastic-EP example \cite{vllm_ep_docs,vllm_elastic_ep_docs}. These systems show that wide EP is becoming practical in serving, but their public abstractions still treat EP membership as control-plane configuration rather than a runtime fault-tolerance problem. They do not address fail-stop loss of an EP rank as an end-to-end problem of safely contracting, restoring expert reachability, and later re-expanding while keeping the instance live. \SystemName is complementary: we implement it in SGLang, but the contribution is the substrate for failure-driven in-place EP repair.

\paragraph{Communication Libraries and EP Backends}

Communication libraries such as NCCL and Gloo provide high-performance collectives over communicator or process-group abstractions whose membership is usually fixed for steady-state execution~\cite{gloo}. Recent NCCL work adds RAS support, communicator shrink and replacement workflows, and EP-oriented APIs in NCCL-EP \cite{nccl_ras,nccl_fault_blog,nccl_ep}. These are important advances, but the central abstraction remains communicator management: when membership changes, the application shrinks or recreates communicators and rebuilds communicator-bound EP state. Our goal is more specific to serving: keep a live EP instance valid across membership changes by making peer membership explicit mutable runtime state.

DeepEP and UCCL-EP move closer to the EP setting \cite{deepep,uccl_ep}. DeepEP focuses on efficient dispatch and combination kernels for a selected EP group, while UCCL-EP targets portable EP communication across hardware and software environments. Both improve EP communication efficiency, but neither addresses the recovery contract we target. In our setting, communication repair alone is insufficient: a rank failure can also remove logical experts and invalidate graph-visible routing state. \SystemName therefore combines peer-set repair, expert-coverage repair, and graph-stable reintegration.

\paragraph{MoE Optimization and Elastic Systems}

Another line of work improves MoE efficiency through routing, kernels, or load balancing, including DeepSpeed-MoE, Tutel, and MegaBlocks \cite{deepspeed_moe,tutel,megablocks}. These systems make MoE faster and more scalable, but do not study how a latency-sensitive EP instance remains correct and available after an unexpected rank failure. More closely related in spirit, serving-oriented systems such as Expert-as-a-Service (EaaS) and ElasticMoE study scalable deployment, autoscaling, and expert placement for MoE services \cite{eaas,elasticmoe}. The key difference is the recovery contract: planned scaling tolerates control-plane orchestration, whereas partial-failure repair must react on the decoding path, quickly re-establish a valid reduced-capacity instance, and later restore capacity without rebuilding the whole instance. Our paper is therefore not about generic elasticity, but about preserving live EP validity under partial faults.

\paragraph{Failure Characterization and Related Systems}

Large-scale studies of ML cluster reliability have characterized failure patterns in production environments. Kokolis et al.~\cite{hpca-reliability} analyze 11 months of data across two 24k-GPU clusters, providing a taxonomy of hardware, system software, and network faults. While their quantitative MTTF figures target training clusters at much larger scales than typical MoE inference deployments, the qualitative insight—that failure probability grows significantly with scale—motivates the need for fault tolerance in wide EP serving. In the training domain, systems like Lazarus~\cite{lazarus} provide elastic fault tolerance for MoE training through adaptive expert replication, while MegaScale-Infer~\cite{megascale-infer} disaggregates attention and FFN modules to improve inference efficiency. Loss-Free Balancing~\cite{loss-free-balancing} offers an auxiliary-loss-free load balancing strategy for MoE training. These works address different points in the design space—training vs. inference, efficiency vs. fault tolerance—and are complementary to \SystemName, which focuses on maintaining live EP validity for inference serving under partial rank failures.
\vspace{-0.3cm}

\section{Conclusion}

Partial-failure tolerance in wide expert parallelism is fundamentally a live EP validity problem. Rather than rebuilding the EP instance after a rank loss, \SystemName repairs the peer set, expert coverage, and routing state that made the instance invalid. Our results show that wide-EP inference need not treat partial failures as global outages: the service can contract, remain useful, and later restore capacity without full reconstruction.

\bibliographystyle{ACM-Reference-Format}
\bibliography{sample-base}

@misc{deepseekai2024deepseekv3technicalreport,
      title={DeepSeek-V3 Technical Report}, 
      author={DeepSeek-AI},
      year={2025},
      eprint={2412.19437},
      archivePrefix={arXiv},
      primaryClass={cs.CL},
      url={https://arxiv.org/abs/2412.19437}, 
}

@misc{tent2026,
      title={TENT: A Declarative Slice Spraying Engine for Performant and Resilient Data Movement in Disaggregated LLM Serving}, 
      author={Feng Ren and Ruoyu Qin and Teng Ma and Shangming Cai and Zheng Liu and Chao Lei and Dejiang Zhu and Ke Yang and Zheming Li and Jialei Cui and Weixiao Huang and Yikai Zhao and Yineng Zhang and Hao Wu and Xiang Gao and Yuhao Fu and Jinlei Jiang and Yongwei Wu and Mingxing Zhang},
      year={2026},
      eprint={2604.00368},
      archivePrefix={arXiv},
      primaryClass={cs.DC},
      url={https://arxiv.org/abs/2604.00368}, 
}

@misc{vllm,
      title={Efficient Memory Management for Large Language Model Serving with PagedAttention}, 
      author={Woosuk Kwon and Zhuohan Li and Siyuan Zhuang and Ying Sheng and Lianmin Zheng and Cody Hao Yu and Joseph E. Gonzalez and Hao Zhang and Ion Stoica},
      year={2023},
      eprint={2309.06180},
      archivePrefix={arXiv},
      primaryClass={cs.LG},
      url={https://arxiv.org/abs/2309.06180}, 
}

@misc{sglang,
      title={SGLang: Efficient Execution of Structured Language Model Programs}, 
      author={Lianmin Zheng and Liangsheng Yin and Zhiqiang Xie and Chuyue Sun and Jeff Huang and Cody Hao Yu and Shiyi Cao and Christos Kozyrakis and Ion Stoica and Joseph E. Gonzalez and Clark Barrett and Ying Sheng},
      year={2024},
      eprint={2312.07104},
      archivePrefix={arXiv},
      primaryClass={cs.AI},
      url={https://arxiv.org/abs/2312.07104}, 
}

@misc{shazeer2017moe,
      title={Outrageously Large Neural Networks: The Sparsely-Gated Mixture-of-Experts Layer}, 
      author={Noam Shazeer and Azalia Mirhoseini and Krzysztof Maziarz and Andy Davis and Quoc Le and Geoffrey Hinton and Jeff Dean},
      year={2017},
      eprint={1701.06538},
      archivePrefix={arXiv},
      primaryClass={cs.LG},
      url={https://arxiv.org/abs/1701.06538}, 
}

@article{switchtransformers,
  author  = {William Fedus and Barret Zoph and Noam Shazeer},
  title   = {Switch Transformers: Scaling to Trillion Parameter Models with Simple and Efficient Sparsity},
  journal = {Journal of Machine Learning Research},
  year    = {2022},
  volume  = {23},
  number  = {120},
  pages   = {1--39},
  url     = {http://jmlr.org/papers/v23/21-0998.html}
}

@misc{vllm_ep_docs,
      title={Expert Parallel Deployment},
      author={{vLLM Team}},
      year={2026},
      howpublished={\url{https://docs.vllm.ai/en/latest/serving/expert_parallel_deployment/}},
      note={vLLM documentation},
}

@misc{vllm_elastic_ep_docs,
      title={Elastic EP},
      author={{vLLM Team}},
      year={2026},
      howpublished={\url{https://docs.vllm.ai/en/latest/examples/online_serving/elastic_ep/}},
      note={vLLM example documentation},
}

@misc{deepep,
      title={DeepEP: an efficient expert-parallel communication library},
      author={Chenggang Zhao and Shangyan Zhou and Liyue Zhang and Chengqi Deng and Zhean Xu and Yuxuan Liu and Kuai Yu and Jiashi Li and Liang Zhao},
      year={2025},
      publisher = {GitHub},
      howpublished = {\url{https://github.com/deepseek-ai/DeepEP}},
}

@misc{gloo,
      title={Gloo: Collective Communications Library},
      author={{PyTorch}},
      year={2026},
      howpublished={\url{https://github.com/pytorch/gloo}},
      note={GitHub repository},
}

@misc{gpudirect_rdma,
      title={GPUDirect RDMA},
      author={{NVIDIA}},
      year={2025},
      howpublished={\url{https://docs.nvidia.com/cuda/gpudirect-rdma}},
      note={CUDA documentation},
}

@misc{cuda_graphs_blog,
      title={Getting Started with CUDA Graphs},
      author={Alan Gray},
      year={2019},
      howpublished={\url{https://developer.nvidia.com/blog/cuda-graphs/}},
      note={NVIDIA Technical Blog},
}

@misc{cuda_programming_guide,
      title={CUDA C Programming Guide},
      author={{NVIDIA}},
      year={2026},
      howpublished={\url{https://docs.nvidia.com/cuda/cuda-programming-guide/}},
      note={CUDA documentation},
}

@misc{nccl_ras,
      title={RAS},
      author={{NVIDIA}},
      year={2025},
      howpublished={\url{https://docs.nvidia.com/deeplearning/nccl/user-guide/docs/troubleshooting/ras.html}},
      note={NCCL documentation},
}

@misc{nccl_fault_blog,
      title={Building Scalable and Fault-Tolerant NCCL Applications},
      author={{NVIDIA}},
      year={2025},
      howpublished={\url{https://developer.nvidia.com/blog/building-scalable-and-fault-tolerant-nccl-applications/}},
      note={NVIDIA Technical Blog},
}

@misc{nccl_ep,
      title={NCCL EP: Towards a Unified Expert Parallel Communication API for NCCL}, 
      author={Amos Goldman and Nimrod Boker and Maayan Sheraizin and Nimrod Admoni and Artem Polyakov and Subhadeep Bhattacharya and Fan Yu and Kai Sun and Georgios Theodorakis and Hsin-Chun Yin and Peter-Jan Gootzen and Aamir Shafi and Assaf Ravid and Salvatore Di Girolamo and Manjunath Gorentla Venkata and Gil Bloch},
      year={2026},
      eprint={2603.13606},
      archivePrefix={arXiv},
      primaryClass={cs.DC},
      url={https://arxiv.org/abs/2603.13606}, 
}

@misc{uccl_ep,
      title={UCCL-EP: Portable Expert-Parallel Communication}, 
      author={Ziming Mao and Yihan Zhang and Chihan Cui and Zhen Huang and Kaichao You and Zhongjie Chen and Zhiying Xu and Zhenyu Gu and Scott Shenker and Costin Raiciu and Yang Zhou and Ion Stoica},
      year={2026},
      eprint={2512.19849},
      archivePrefix={arXiv},
      primaryClass={cs.DC},
      url={https://arxiv.org/abs/2512.19849}, 
}

@misc{deepspeed_moe,
      title={DeepSpeed-MoE: Advancing Mixture-of-Experts Inference and Training to Power Next-Generation AI Scale}, 
      author={Samyam Rajbhandari and Conglong Li and Zhewei Yao and Minjia Zhang and Reza Yazdani Aminabadi and Ammar Ahmad Awan and Jeff Rasley and Yuxiong He},
      year={2022},
      eprint={2201.05596},
      archivePrefix={arXiv},
      primaryClass={cs.LG},
      url={https://arxiv.org/abs/2201.05596}, 
}

@misc{tutel,
      title={Tutel: Adaptive Mixture-of-Experts at Scale}, 
      author={Changho Hwang and Wei Cui and Yifan Xiong and Ziyue Yang and Ze Liu and Han Hu and Zilong Wang and Rafael Salas and Jithin Jose and Prabhat Ram and Joe Chau and Peng Cheng and Fan Yang and Mao Yang and Yongqiang Xiong},
      year={2023},
      eprint={2206.03382},
      archivePrefix={arXiv},
      primaryClass={cs.DC},
      url={https://arxiv.org/abs/2206.03382}, 
}

@misc{megablocks,
      title={MegaBlocks: Efficient Sparse Training with Mixture-of-Experts}, 
      author={Trevor Gale and Deepak Narayanan and Cliff Young and Matei Zaharia},
      year={2022},
      eprint={2211.15841},
      archivePrefix={arXiv},
      primaryClass={cs.LG},
      url={https://arxiv.org/abs/2211.15841}, 
}

@misc{eaas,
      title={Expert-as-a-Service: Towards Efficient, Scalable, and Robust Large-scale MoE Serving}, 
      author={Ziming Liu and Boyu Tian and Guoteng Wang and Zhen Jiang and Peng Sun and Zhenhua Han and Tian Tang and Xiaohe Hu and Yanmin Jia and Yan Zhang and He Liu and Mingjun Zhang and Yiqi Zhang and Qiaoling Chen and Shenggan Cheng and Mingyu Gao and Yang You and Siyuan Feng},
      year={2025},
      eprint={2509.17863},
      archivePrefix={arXiv},
      primaryClass={cs.DC},
      url={https://arxiv.org/abs/2509.17863}, 
}

@misc{elasticmoe,
      title={ElasticMoE: An Efficient Auto Scaling Method for Mixture-of-Experts Models}, 
      author={Gursimran Singh and Timothy Yu and Haley Li and Cheng Chen and Hanieh Sadri and Qintao Zhang and Yu Zhang and Ying Xiong and Yong Zhang and Zhenan Fan},
      year={2025},
      eprint={2510.02613},
      archivePrefix={arXiv},
      primaryClass={cs.DC},
      url={https://arxiv.org/abs/2510.02613}, 
}

@misc{deepseekmoe,
      title={DeepSeekMoE: Towards Ultimate Expert Specialization in Mixture-of-Experts Language Models},
      author={Damai Dai and Chengqi Deng and Chenggang Zhao and Runxin Xu and Huazuo Gao and Deli Chen and Jiashi Li and Wangding Zeng and Xingkai Yu and Yu Wu and Zhenda Xie and Y.K. Li and Panpan Huang and Fuli Luo and Chong Ruan and Zhifang Sui and Wenfeng Liang},
      year={2024},
      eprint={2401.06066},
      archivePrefix={arXiv},
      primaryClass={cs.CL},
      url={https://arxiv.org/abs/2401.06066},
}

@misc{deepseekv2,
      title={DeepSeek-V2: A Strong, Economical, and Efficient Mixture-of-Experts Language Model},
      author={DeepSeek-AI and Qihao Zhu and Daya Guo and Zhihong Shao and Dejian Yang and Peiyi Wang and Runxin Xu and Y. Wu and Yukun Li and Huazuo Gao and Shirong Ma and Wangding Zeng and Xiao Bi and Zihui Gu and Hanwei Xu and Damai Dai and Kai Dong and Liyue Zhang and Yishi Piao and Zhibin Gou and Zhenda Xie and Zhewen Hao and Bing-Li Wang and Jun-Mei Song and Deli Chen and Xin Xie and Kang Guan and Yu mei You and Aixin Liu and Qiushi Du and Wenjun Gao and Xuan Lu and Qinyu Chen and Yaohui Wang and Chengqi Deng and Jiashi Li and Chenggang Zhao and Chong Ruan and Fuli Luo and Wenfeng Liang},
      year={2024},
      eprint={2405.04434},
      archivePrefix={arXiv},
      primaryClass={cs.CL},
      url={https://arxiv.org/abs/2405.04434},
}

@misc{nvshmem-ibgda,
      title={Improving Network Performance of {HPC} Systems Using {NVIDIA} Magnum {IO} {NVSHMEM} and {GPUDirect} Async},
      author={Pak Markthub and Jim Dinan and Sreeram Potluri and Seth Howell},
      year={2022},
      howpublished={NVIDIA Technical Blog},
      url={https://developer.nvidia.com/blog/improving-network-performance-of-hpc-systems-using-nvidia-magnum-io-nvshmem-and-gpudirect-async/},
}

@misc{loss-free-balancing,
      title={Auxiliary-Loss-Free Load Balancing Strategy for Mixture-of-Experts},
      author={Lean Wang and Huazuo Gao and Chenggang Zhao and Xu Sun and Damai Dai},
      year={2024},
      eprint={2408.15664},
      archivePrefix={arXiv},
      primaryClass={cs.LG},
      url={https://arxiv.org/abs/2408.15664},
}

@misc{nccl-gin,
      title={GPU-Initiated Networking for {NCCL}},
      author={Khaled Hamidouche and John Bachan and Pak Markthub and Peter-Jan Gootzen and Elena Agostini and Sylvain Jeaugey and Aamir Shafi and Georgios Theodorakis and Manjunath Gorentla Venkata},
      year={2025},
      eprint={2511.15076},
      archivePrefix={arXiv},
      primaryClass={cs.DC},
      url={https://arxiv.org/abs/2511.15076},
}

@misc{eplb,
      title={Expert Parallelism Load Balancer ({EPLB})},
      author={DeepSeek-AI},
      year={2025},
      howpublished={GitHub},
      url={https://github.com/deepseek-ai/EPLB},
}

@misc{megascale-infer,
      title={MegaScale-Infer: Serving Mixture-of-Experts at Scale with Disaggregated Expert Parallelism},
      author={Ruidong Zhu and Ziheng Jiang and Chao Jin and Peng Wu and Cesar A. Stuardo and Dongyang Wang and Xinlei Zhang and Huaping Zhou and Haoran Wei and Yang Cheng and Jianzhe Xiao and Xinyi Zhang and Lingjun Liu and Haibin Lin and Li-Wen Chang and Jianxi Ye and Xiao Yu and Xuanzhe Liu and Xin Jin and Xin Liu},
      year={2025},
      eprint={2504.02263},
      archivePrefix={arXiv},
      primaryClass={cs.DC},
      url={https://arxiv.org/abs/2504.02263},
}

@misc{lazarus,
      title={Lazarus: Resilient and Elastic Training of Mixture-of-Experts Models with Adaptive Expert Placement},
      author={Yongji Wu and Wenjie Qu and Xueshen Liu and Tianyang Tao and Yifan Qiao and Zhuang Wang and Wei Bai and Yuan Tian and Jiaheng Zhang and Z. Morley Mao and Matthew Lentz and Danyang Zhuo and Ion Stoica},
      year={2024},
      eprint={2407.04656},
      archivePrefix={arXiv},
      primaryClass={cs.DC},
      url={https://arxiv.org/abs/2407.04656},
}

@inproceedings{hpca-reliability,
      title={Revisiting Reliability in Large-Scale Machine Learning Research Clusters},
      author={Apostolos Kokolis and Michael Kuchnik and John Hoffman and Adithya Kumar and Parth Malani and Faye Ma and Zachary DeVito and Shubho Sengupta and Kalyan Saladi and Carole-Jean Wu},
      booktitle={2025 IEEE International Symposium on High-Performance Computer Architecture (HPCA)},
      year={2025},
      publisher={IEEE},
}

\end{document}